\documentclass[12pt,a4paper]{article}

\usepackage{algorithm}
\usepackage{algpseudocode}
\usepackage{geometry}
\RequirePackage{amsmath,amssymb}
\RequirePackage{mathrsfs}
\RequirePackage{amsfonts,bm}
\RequirePackage{dsfont}
\RequirePackage{braket}
\RequirePackage{tabularx}
\usepackage{multirow}
\RequirePackage{nicefrac}
\RequirePackage{graphicx}
\RequirePackage{booktabs}
\RequirePackage{colortbl}
\RequirePackage{xcolor}
\RequirePackage[symbol]{footmisc}
\usepackage{mathtools}
\usepackage{comment}
\usepackage{blkarray}
\usepackage{stackengine}
\usepackage{float}
\usepackage{rotating}
\usepackage{hhline}
\usepackage{tikz}
\usepackage{multirow}
\usepackage[section]{placeins}

\def\AEF{A.E. Faraggi}

\def\IJMP#1#2#3{{\it Int.\ J.\ Mod.\ Phys.}\/ {\bf A#1} (#2) #3}

\def\EJP#1#2#3{{\it Eur.\ Phys.\ Jour.}\/ {\bf C#1} (#2) #3}

\def\JHEP#1#2#3{{\it JHEP}\/ {\bf #1} (#2) #3}
\def\NPB#1#2#3{{\it Nucl.\ Phys.}\/ {\bf B#1} (#2) #3}
\def\PLB#1#2#3{{\it Phys.\ Lett.}\/ {\bf B#1} (#2) #3}
\def\PRD#1#2#3{{\it Phys.\ Rev.}\/ {\bf D#1} (#2) #3}

\def\PRT#1#2#3{{\it Phys.\ Rep.}\/ {\bf#1} (#2) #3}
\def\etal{{\it et al\/}}

\def\beq{\begin{equation}}
\def\eeq{\end{equation}}
\def\beqn{\begin{eqnarray}}
\def\eeqn{\end{eqnarray}}
\def\ds{{$\tilde S$}}
\def\unahe{{${\overline{\rm NAHE}}$}}
\usepackage{empheq}
\usepackage[most]{tcolorbox}
\usepackage{blkarray}
\newtcbox{\mymath}[1][]{%
    nobeforeafter, math upper, tcbox raise base,
    enhanced, colframe=blue!30!black,
    colback=blue!30, boxrule=1pt,
    #1}

\newcommand{\cc}[2]{c{#1\atopwithdelims[]#2}}

\newcommand{\nn}{\nonumber}
\newcommand{\ba}{\begin{eqnarray}}
\newcommand{\ea}{\end{eqnarray}}

\numberwithin{equation}{section}

\begin{document}
\begin{titlepage}
\samepage{
\setcounter{page}{1}
\rightline{LTH-1306}
\rightline{June 2022}

\vfill
\begin{center}
  {\Large
    \bf{Spinor--Vector Duality and the Swampland
    }
  }

\vspace{1cm}
\vfill

{\large
  Alon E. Faraggi$^{1,2}$\footnote{E-mail address: alon.faraggi@liverpool.ac.uk}
}

\vspace{1cm}

{\it $^{1}$ Dept.\ of Mathematical Sciences, University of Liverpool, Liverpool
L69 7ZL, UK\\}
\vspace{.08in}

{\it $^{2}$ Dept. of Particle Physics and Astrophysics, Weizmann Institute,
Rehovot 76100, Israel\\}

\vspace{.025in}
\end{center}

\vfill
\begin{abstract}
\noindent
The Swampland Program aims to address the question, "when does an effective field theory model
of quantum gravity have an ultraviolet complete embedding in string theory?", and can be regarded
as a bottom--up approach to investigations of quantum gravity. An alternative top--down approach aim
to explore the imprints and the constraints imposed by the string theory dualities and symmetries 
on the effective field theory representations of quantum gravity. The most celebrated example 
of this approach is mirror symmetry. Mirror symmetry was first observed in worldsheet 
contructions of string compactifications. It was completely unexpected from the effective 
field theory point of view, and its implications in that context were astounding. In terms of 
the moduli parameters of toroidally compactified Narain spaces, mirror symmetry can be regarded as 
arising from mappings of the moduli of the internal compactified space. Spinor--vector duality, which
was discovered in worldsheet constructions of string vacua, is an extension of mirror symmetry 
that arises from mappings of the Wilson line moduli, and provide a probe to constrain and explore 
the moduli spaces of $(2,0)$ string compactifications. Mirror symmetry and spinor--vector duality are mere
two examples of a much wider symmetry structure, whose implications are yet to be unravelled.
A mapping between supersymmetric and non--supersymmetric vacua is briefly discussed. 
$T$--duality is another important property 
of string theory, and can be thought of as phase--space duality in compact space. I propose that
manifest phase--space duality, and the related equivalence postulate of quantum mechanics, 
provide the background independent overarching principles underlying quantum gravity. 
\end{abstract}

\smallskip}

\end{titlepage}

\section{Introduction}

Physics is first and foremost an experimental science. Be that as it may, the language which is used to encode experimental observations is mathematics. It therefore makes sense to construct mathematical models that aim to describe physical reality, where successful mathematical models are those that are able to account for a wider range of observational data. Over the past century the mathematical modelling of physical observations cumulated in two fundamental theories. In the subatomic domain quantum mechanics and its incarnation in the form of Quantum Field Theory (QFT) accounts for all the available data, whereas in the celestial, galactical and cosmological realms Einstein's general relativity theory of gravity is used to parametrise the observations. Yet, these two basic theories are fundamentally incompatible. The incompatibility is particularly glaring when it comes to the vacuum. Whereas gravitationally based observations mandate that the vacuum energy is very small, quantum field theory models that are used to describe the subatomic data predict vacuum energy that by far exceed the physical observations. There are further basic incompatibilities between the two theories that are to do with their distinct mathematical formulations, {\it e.g.} the black hole information paradox and the unrenormalisability of quantum field theory formulations of general relativity. 

Attempts to construct consistent mathematical formulations of quantum gravity therefore occupy and motivate much of the research in fundamental physics. There are numerous approaches to this problem that include 
string theory \cite{stringtheory} and loop quantum gravity \cite{loopy}. String theory is a mundane extension of point particle quantum mechanics that provides a perturbatively self consistent framework for quantum gravity. The consistency requirements of string theory mandate the existence of the gauge and matter structures that are the bedrock of the Standard Model of particle physics. String theory therefore provides an effective framework to develop a phenomenological approach to quantum gravity. The consistency
conditions of string theory require additional degrees of freedom
beyond those observed in the Standard Models of particle physics and
cosmology. In some guises these can be interepreted as additional
bosonic spacetime dimensions, with 26 extra dimensions required
in the bosonic string and 10 in the fermionic, whereas the
heterotic--string is a hybrid of a left--moving fermionic sector and
right--moving bosonic sector. 

The number of string theories in higher dimensions is relatively scarce, and
includes five theories that are supersymmetric and eight that
are not. Moreover, these ten dimensional theories are 
connected in a lower dimension by perturbative
interpolations or orbifolds, or by some non--perturbative
transformations. The extra dimensions are compactified
on an internal space such that they are hidden from contemporary
experimental observations, resulting in a plethora of vacua
in lower dimensions. Nevertheless, there may exist symmetries
that underlie the entire space of vacua in lower dimensions,
akin, perhaps, to the symmetries that underlie the vacua in
higher dimensions. Mirror symmetry is an example of such a symmetry
between vacua in lower dimensions \cite{mirrorsymmetry, mirrorreview}.

String vacua in four dimensions are in general studied by using
exact worldsheet constructions, as well as effective field theory
target space tools that explore the low energy spectrum
of string compactifications. Among the exact worldsheet
tools we may list the toroidal orbifolds \cite{dhvw},
the fermionic formulation \cite{fff}, and the
interacting Conformal Field Theory constructions \cite{gepner}.
The effective field theory models typically are obtained as
compactifications of ten or eleven dimensional supergravity
on a complex Ricci flat internal manifold \cite{candelas, IbUranga}. 

A fundamental issue in this regard
is the relation between the exact string solutions and their
low energy effective field theory and vise versa. 
This question motivates much of the contemporary interest in string phenomenology, 
in the context of the "so called" swampland program \cite{swamppro},
which aims to address the question when can
an effective field theory model of quantum gravity be completed to a fully fledged string theory
and when it cannot. The swampland program further aims to uncover the fundamental principles that 
underlie quantum gravity. The swampland program \cite{swamprev}
can be viewed as a bottom--up approach to 
the exploration of the synthesis of quantum mechanics with gravity. 

These questions are vital because at the end of the day it is likely
that the correlation of a string vacuum with experimental data is
performed by using its effective field theory limit, whereas the
string construction may produce the boundary data in the form of the
gauge and Yukawa couplings. To date the relation between exact
string solutions and their effective field theory smooth limit
is only well understood in limited cases \cite{DKL}, and 
entail mostly the analysis of various supergravity theories that are
EFT limits of the corresponding string theories.

An alternative approach to the swampland program is a top--down approach that aims to explore
the symmetries of exact string solutions and their imprints in the effective field theory limit.
The questions then are two fold: 1. what is the complete set of symmetries that underlie the moduli spaces of string vacua; 2. can we guess from this set of symmetries the general principles that underlie quantum gravity. 

Mirror symmetry is an example of a symmetry that was observed
initially via the exact worldsheet Conformal Field Theory (CFT) constructions with
profound implications for the geometrical spaces that are
utilised in the effective field theory limits \cite{canxen}.
The exact worldsheet string theories have a rich symmetry structure
that arises due to the exchange of massless and massive modes.
Mirror symmetry is believed to be related to T--duality \cite{syz}.
In toroidal orbifold compactifications T--duality arises
due to the exchange of the moduli of the internal six
dimensional compactified manifold \cite{gprreview}.
The toroidal orbifold compactifications of the
heterotic--string have additional moduli that correspond to
Wilson line moduli. 

Spinor--Vector Duality (SVD) \cite{svd1,svd2}
is a map between dual string vacua that can be understood as a result of exchanging Wilson line moduli \cite{ffmt}. In this sense the spinor--vector duality is an extension of mirror symmetry. The SVD operates under the exchange of the total number
of spinorial plus anti--spinorial representations and the total number
of vectorial representations of the underlying GUT symmetry group,
where the GUT group is $SO(12)$, or $SO(10)$, in the case of models with
$N =2$, or $N=1$, spacetime supersymmetry, respectively.
The spinor--vector duality which is observed in $Z_2\times Z_2$ orbifold
compactifications generalises to exact string solutions with interacting
internal CFT \cite{afg}.

Like mirror symmetry we can seek to explore the imprint of SVD in the effective field theory limit of string compactifications. This program was initiated over the past couple of years by studying resolutions of dual orbifold models. In $Z_2\times Z_2$ orbifolds SVD is realised plane by plane, and hence arises in vacua with a single $Z_2$ of the internal coordinates, whereas a second $Z_2$ action corresponds to the Wilson line \cite{fkrneq2}. These cases were studied in the effective field theory limits in 5 \cite{5dsvd} and 6 dimensions \cite{6dsvd}. The SVD in the full $Z_2\times Z_2$ orbifold requires some refinement of the orbifold singularity resolution tools that was developed in \cite{taming}. The exploration of SVD in the effective field theory limit entails the analysis of complex manifolds with vector bundles. The SVD therefore provides a useful tool to explore and constrain the moduli spaces of Calabi-Yau manifolds with vector bundles. The next question that one may entertain is what is the complete set of symmetries that underlie the string vacua and whether symmetries such as mirror symmetry, or SVD provide such a complete set, {\it i.e.} does every string vacuum have a mirror dual; does every $Z_2$ string vacuum with a number of spinorial plus anti--spinorial respresentations and vectorial representations of the GUT group have a dual vacuum in which the numbers are interchanged. We may conjecture that this is indeed the case and that an effective field theory model that does not have such a dual is necessarily in the swampland. 

We may envision that mirror symmetry and SVD are mere reflections of a much wider symmetry structure that underlies the space of string vacua. This much wider space can be glimpsed by compactifying the string models to 2 dimensions. Compactifications to 2 dimensions give rise to 24 dimensional lattices that exhibit a very rich symmetry structure. We may explore the possibility that mirror symmetry and the SVD are reflections of this wider symmetry structure and whether this symmetry structure corresponds to a finite and complete space. The fact that string theory predicts that a finite number of degrees of freedom is required to obtain a perturbatively consistent theory of quantum gravity gives hope that this is indeed the case. %Quantum gravity is solvable. 

We have learned that SVD is a mere extension of the duality symmetries that underlie the space of string vacua under exchanges of Wilson line moduli. In the context of toroidal compactifications the full set of symmetries correspond to exchanges of the parameters of the background fields that include the metric, the anti-symmetric tensor field and the Wilson line moduli. Exchanges of the parameters of the metric and antisymmetric tensor field correspond to generalised $T$--duality symmetries \cite{gprreview}. We may interpret $T$--duality as phase--space duality in compact space, {\it i.e.} $T$--duality exchanges momentum modes with winding modes that we may regard as the phase--space of the compact space. Requiring manifest phase--space duality has been the starting point in the development of the Equivalence Postulate approach to Quantum Mechanics. In the EPOQM, quantum mechanics is derived from a geometrical principle. We may speculate that the requirement of manifest phase--space duality and the equivalence postulate of quantum mechanics
provide the overarching principles that underlie quantum gravity.
%
%In this short essay I will describe the main ingredients of the symmetry structures alluded to above. 

\section{SVD in four dimensional worldsheet constructions }
  
The spinor--vector duality was first observed in free fermionic constructions of the heterotic--string in four dimensions. 
The SVD was initially observed by simple counting \cite{svd1}, using the
classification tools developed in \cite{gkr} for type II string, and
in \cite{fknr, svd1, svd2}
for heterotic--strings with unbroken $SO(10)$ GUT symmetry and 
$N=1$ spacetime supersymmetry.

%\begin{center}
%\begin{tabular}{|ccc|ccc|ccc|c|}
%\hline
%      & First Plane & &    & Second plane & & & Third Plane & & \\
%\hline
%$s$ &${\bar s}$& $v$ &$s$&${\bar s}$&$v$&$s$&${\bar s}$&$v$& \# of models \\
%\hline
%2 & 0 & 0 &    0 & 0 & 0 &    0 & 0 & 0 & 1325963712 \\
%0 & 2 & 0 &    0 & 0 & 0 &    0 & 0 & 0 & 1340075584 \\
%1 & 1 & 0 &    0 & 0 & 0 &    0 & 0 & 0 & 3718991872 \\
%\hline
%0 & 0 & 2 &    0 & 0 & 0 &    0 & 0 & 0 & 6385031168 \\
%\hline
%\hline
%\end{tabular}
%\label{svdcounting}
%\end{center}

\begin{table}[H] 
\setlength{\tabcolsep}{5mm}
%\begin{center}
\begin{tabularx}{\textwidth}{ccc|ccc|ccc|c}
\toprule
 \multicolumn{3}{c}{\textbf{First Plane}} &  \multicolumn{3}{c}{ \textbf{Second Plane}} &    \multicolumn{3}{c}{\textbf{Third Plane }}&  \\
\midrule
\boldmath{$s$} &\boldmath{${\bar s}$}& \boldmath{$v$} &\boldmath{$s$}&\boldmath{${\bar s}$}&\boldmath{$v$}&\boldmath{$s$}&\boldmath{${\bar s}$}&\boldmath{$v$}& \textbf{\# of Models} \\
\midrule
2 & 0 & 0 &    0 & 0 & 0 &    0 & 0 & 0 & 1325963712 \\
0 & 2 & 0 &    0 & 0 & 0 &    0 & 0 & 0 & 1340075584 \\
1 & 1 & 0 &    0 & 0 & 0 &    0 & 0 & 0 & 3718991872 \\
\midrule
0 & 0 & 2 &    0 & 0 & 0 &    0 & 0 & 0 & 6385031168 \\
\bottomrule
\end{tabularx}
\caption{Number of models with a total number of 2 representations in the first plane. 
}
\label{svdcounting}
\end{table}
 
This is illustrated in table \ref{svdcounting}, 
\noindent
This is illustrated in the table above, where $s$, ${\bar s}$ and $v$ 
refer to the total number of spinorial $16$, anti--spinorial $\overline{16}$, and vectorial
$10$ representations, respectively. %\ref{svdcounting}. 
The SVD operates plane by plane of the $Z_2\times Z_2$ orbifold. It is readily 
checked that the total number of models with 2 spinorial, 2 anti--spinorial, and 1 spinorial plus 1 anti--spinorial, representations, of the $SO(10)$ GUT group, is the same as the total
number of models with 2 vectorial representations. This is easily checked by adding the numbers in the first three rows, which is equal to the number in the last row. 

The SVD was subsequently proven to arise due to exchange of discrete GGSO
phases in the free fermionic formulation \cite{svd2, cfkr}, or
due to discrete torsions in a toroidal orbifold representation
\cite{aft, ffmt}. The spinor--vector duality is readily understood
if we consider the case in which an $SO(10)\times U(1)$ symmetry
is enhanced to $E_6$. In this case the string compactification
possesses a (2,2) worldsheet supersymmetry. The representation of
$E_6$ are the chiral $27$ and and anti--chiral 
$\overline{27}$,
which decompose as $27=16_{+1/2}+10_{-1}+1_{+2}$
and $\overline{27}=\overline{16}_{-1/2}+10_{+1}+1_{-2}$
under $SO(10)\times U(1)$.

If one now counts the total number of $\#_1 (16+\overline{16})$
and $\#_2 (10)$, it is apparent that in this case $\#_1= \#_2$.
That is, the point in the moduli space in which the symmetry
is enhanced to $E_6$ is a self--dual point under SVD. This is
similar to the case of T--duality on a circle, in which at
the self--dual point under T--duality the gauge symmetry is
enhanced from $U(1)$ to $SU(2)$. Away from the self--dual point
the $E_6$ symmetry is broken to $SO(10)\times U(1)$ and the
worldsheet supersymmetry is broken from $(2,2)$ to $(2,0)$.
In general the $E_6$ symmetry is broken in the toroidal
orbifold models by Wilson lines, or by some discrete phases,
whereas in the fermionic
language they may appear as Generalised GGSO phases,
in the one--loop partition function.
The SVD duality states that for any string vacuum, in which
$E_6\rightarrow SO(10)\times U(1)$, with $\#_1(16+\overline{16})$
and $\#_2(10)$ representations, there exist a dual vacuum
in which $\#_1\leftrightarrow\#_2$. 

\begin{figure}[h]
\includegraphics[width=14pc]{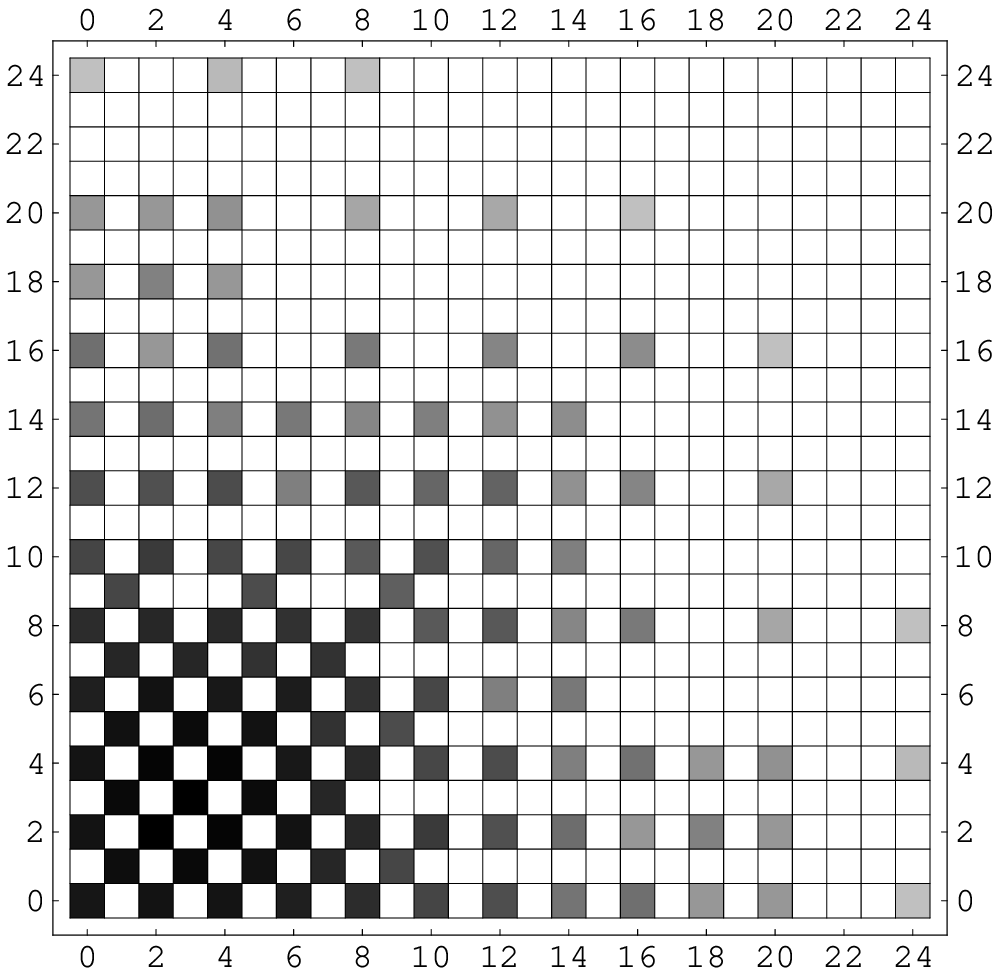}\hspace{2pc}%
\begin{minipage}[b]{21pc}
\caption{
\label{den}
Density plot showing the spinor--vector duality in the space of fermionic
$Z_2\times Z_2$ heterotic--string models. The plot shows the number 
of vacua with a given number of $({16}+\overline{16})$ and 
{10} multiplets of $SO(10)$. It is invariant under exchange of 
rows and columns, reflecting the spinor--vector duality 
underlying the entire space of vacua. Models on the diagonal are
self--dual under the exchange of rows and columns, {\it i.e.}
$\# ({16}+\overline{16}) = \#({10})$ without enhancement to
$E_6$, which are self--dual by virtue of the enhanced symmetry. 
}
\end{minipage}
\end{figure}

The 
spinor--Vector Duality (SVD) is depicted in figure \ref{den}, 
which shows a distribution of the number of models with a 
$\#_1 (16+\overline{16})$ and $\#_2 (10)$. Figure \ref{den} 
is symmetric under exchange of rows and columns reflecting the 
duality under the spinor--vector exchange. 

As the free fermionic 
heterotic--string vacua correspond to the $Z_2\times Z_2$
orbifolds, they contain three twisted sectors that preserve each
an $N=2$ spacetime supersymmetry. The SVD is realised in fact
in each twisted sector separately, {\it i.e.} it can be realised
in models that possess $N=2$, rather than $N=1$, spacetime
supersymmetry \cite{fkrneq2}.
In the $N=2$ vacua the enhanced symmetry at the self--dual
point is $E_7$, which is broken to $SO(12)\times SU(2)$ away from
the self--dual point, and the SVD is realised in terms of the relevant
representations of $E_7$ and $SO(12)\times SU(2)$ \cite{fkrneq2}.

\begin{figure}[!]
	\centering
	\includegraphics[width=100mm]{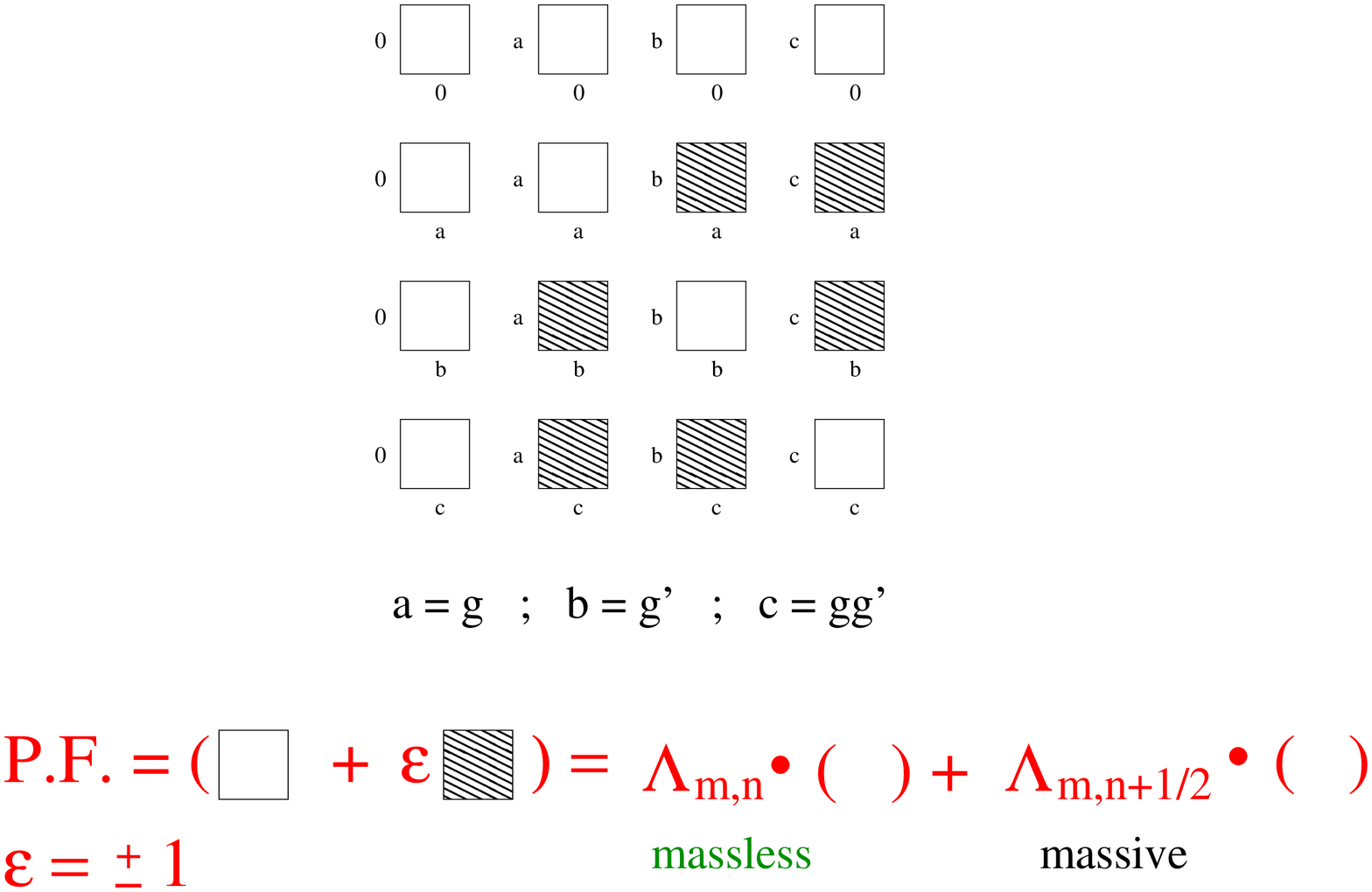}
	\caption{{
            The
            $Z_2\times Z_2^\prime$ partition function of the $g^\prime$--twist and $g$ Wilson line
            with discrete torsion $\epsilon=\pm1$. 
}
}
	\label{z2z2svd}
\end{figure}

Further understanding of the spinor--vector duality is gained
by translating to the bosonic $Z_2\times Z_2$ representation.
Since, the SVD operates in each of the $Z_2$ planes separately, 
we can study it in vacua with a single $Z_2$ twist of the
compactified coordinates \cite{fkrneq2}.
Using the level one $SO(2n)$ characters \cite{angelsign},
\beqn
& & O_{2n} = {1\over 2} \left( {\theta_3^n \over \eta^n} +
{\theta_4^n \over \eta^n}\right) ~~~~~~\,,
~~~~~~~~~~~~
V_{2n} = {1\over 2} \left( {\theta_3^n \over \eta^n} -
{\theta_4^n \over \eta^n}\right) \,,
~~~~~~~~~~~~
\label{so2ncharaOV} \\
& & S_{2n} = {1\over 2} \left( {\theta_2^n \over \eta^n} +
i^{-n} {\theta_1^n \over \eta^n} \right) ~~\,,
~~~~~~~
C_{2n} = {1\over 2} \left( {\theta_2^n \over \eta^n} -
i^{-n} {\theta_1^n \over \eta^n} \right) \,,
\label{so2ncharaSC}
\eeqn
{where} 
\beq
{
\theta_3\equiv Z_f{0\choose0}~~~,~~~
  \theta_4\equiv Z_f{0\choose1}~~~},~~~
{
  \theta_2\equiv Z_f{1\choose0}~~~,~~~
  \theta_1\equiv Z_f{1\choose1}~,~~}\nonumber
\eeq
and $Z_f$ is the partition function of a single
complex worldsheet fermion, in terms of 
theta functions.
The partition function
of the $E_8\times E_8$  heterotic--string reduced to four
dimensions is given by
\beq
{Z}_+ = (V_8 - S_8) \, 
\left( \sum_{m,n} \Lambda_{m,n}\right)^{\otimes 6}\, 
\left(\overline{ O} _{16} + \overline{ S}_{16} \right) 
\left(\overline{O} _{16} + \overline{ S}_{16} \right)\,,
\label{zplus}
\eeq
where for each $S_1$,
\beq
p_{\rm L,R}^i = {m_i \over R_i} \pm {n_i R_i \over \alpha '} \,
~~~~~~~~~
{\rm and}
~~~~~~~~~
\Lambda_{m,n} = {q^{{\alpha ' \over 4} 
p_{\rm L}^2} \, \bar q ^{{\alpha ' \over 4} 
p_{\rm R}^2} \over |\eta|^2}\,.
\nonumber
\eeq
A $Z_2\times Z_2^\prime:g\times g^\prime$ action on $Z_+$ is applied.
The first $Z_2$ couples a fermion number in the observable and hidden sectors with 
a $Z_2$--shift in a compactified coordinate, and is given by
$
g: (-1)^{(F_{1}+F_2)}\delta
$
~where the fermion numbers $F_{1,2}$ operate on the spinorial
representations of the observable and hidden $SO(16)$ groups as
$$
F_{1,2}:({\overline O}_{16}^{1,2},
             {\overline V}_{16}^{1,2},
             {\overline S}_{16}^{1,2},
             {\overline C}_{16}^{1,2})\longrightarrow~
            ({\overline O}_{16}^{1,2},
             {\overline V}_{16}^{1,2},
             -{\overline S}_{16}^{1,2},
             -{\overline C}_{16}^{1,2})
$$
~and $\delta$ identifies points shifted by a $Z_2$ shift 
in the $X_9$ direction, {\it i.e.} 
$
\delta X_9 = X_9 +\pi R_9.~
$
The effect of the shift is to insert a factor of $(-1)^m$ into the lattice 
sum in eq. (\ref{zplus}), {\it i.e.} 
$
\delta:\Lambda_{m,n}^9\longrightarrow(-1)^m\Lambda_{m,n}^9.
$
~The second $Z_2$ operates as a twist on the internal coordinates 
given by 
\beq
{g^\prime}:(x_{4},x_{5},x_{6},x_7,x_8,x_9)
\longrightarrow
(-x_{4},-x_{5},-x_{6},-x_7,+x_8,+x_9). 
\label{z2twist}
\eeq
Alternatively, the first $Z_2$ action can be interpreted as a Wilson line
in $X_9$ \cite{ffmt},
$$
g~: (0^7,1|1, 0^7)  ~\rightarrow~
E_8\times E_8\rightarrow SO(16)\times SO(16).\nonumber
$$
The $Z_2$ space twisting breaks $N=4\rightarrow N=2$ spacetime
supersymmetry and $E_8\rightarrow E_7\times SU(2)$,
or with the inclusion of the Wilson line $SO(16)\rightarrow SO(12)\times SO(4)$.
The orbifold partition function is
$${Z~=~
\left({Z_+\over{Z_g\times Z_{g^{\prime}}}}\right)~=~
\left[{{(1+g)}\over2}{{(1+g^\prime)}\over2}\right]~Z_+}.$$
The partition function includes an untwisted sector and three twisted sectors.
Its schematic form is shown in figure \ref{z2z2svd}.

The winding modes in the sectors twisted by 
$g$ and $gg^\prime$ are shifted by $1/2$. These sectors therefore only
contain massive states. The sector twisted by $g^\prime$
produces massless twisted matter states. 
The partition function has two modular orbits
and one discrete torsion $\epsilon=\pm1$ between the two orbits. 
Massless states are obtained for zero lattice modes. 
The terms in the sector $g^\prime$ contributing to the massless 
spectrum take the form
\beqn
& &     \Lambda_{p,q}
\left\{
 {1\over2}
\left( 
           \left\vert{{2\eta}\over\theta_4}\right\vert^4
         +
           \left\vert{{2\eta}\over\theta_3}\right\vert^4
\right)
\left[{
       P_\epsilon^+Q_s{\overline V}_{12}{\overline C}_4{\overline O}_{16}} +
   {P_\epsilon^-Q_s{\overline S}_{12}{\overline O}_4{\overline O}_{16} }
\right.
{\left.  \right] + }
\right. \nonumber\\ 
& &\nonumber\\ 
& &\left.
~~~~~~~~~~{1\over2}\left(    \left\vert{{2\eta}\over\theta_4}\right\vert^4
                      -
                         \left\vert{{2\eta}\over\theta_3}\right\vert^4\right)
\left[{
P_\epsilon^+Q_s
{\overline O}_{12}{\overline S}_4{\overline O}_{16}} \right.
{\left. \right] } 
\right\}~~~~~
+~~\hbox{massive}
 \label{masslessterminpf}
\eeqn
where 
\beq
P_\epsilon^+~=~\left({{1+\epsilon(-1)^m}\over2}\right)\Lambda_{m,n}~~~;~~~
P_\epsilon^-=\left({{1-\epsilon(-1)^m}\over2}\right)\Lambda_{m,n} 
\label{pepluspeminus}
\eeq
Depending on the sign of $\epsilon=\pm$ 
it is noted from  eq. (\ref{pepluspeminus}) that either the spinorial states, 
or the vectorial states, are massless. In the case with $\epsilon=+1$ 
we see from eq. (\ref{eplus}) that in this case massless momentum
modes from the shifted lattice arise in $P_\epsilon^+$ whereas 
$P_\epsilon^-$ gives rise to massive modes. Therefore, in this case  
the vectorial character ${\overline V}_{12}$ in eq. (\ref{pepluspeminus})
produces massless states, whereas the spinorial character
${\overline S}_{12}$ generates massive states.
In the case with $\epsilon=-1$ eq. (\ref{eminus}) reveals
that exactly the opposite occurs. 
\beqn
{\epsilon~=~+1~~}&{\Rightarrow}&
{~~P^+_\epsilon~=~~~~~~~~~~~\Lambda_{2m,n}~~~
~~~~~~~~~~~~{ P^-_\epsilon~=~~~~~~~~
\Lambda_{2m+1,n}}~~~}\label{eplus}\\
{\epsilon~=~-1~~}&{\Rightarrow}&
{{ ~~P^+_\epsilon~=~~~~~~~~~~~\Lambda_{2m+1,n}}~~~
~~~~~~~~~P^-_\epsilon~=~~~~~~~~\Lambda_{2m,n}~~~}\label{eminus}
\eeqn
It is noted that 
the spinor--vector duality is produced by the exchange of the discrete torsion
$\epsilon=+1\rightarrow \epsilon =-1$ in the $Z_2\times Z_2^\prime$ partition function.
This is similar to the case of mirror symmetry in the $Z_2\times Z_2$
orbifold of ref. \cite{vafawitten}, where the mirror symmetry map is generated
by the exchange of the discrete torsion between the two $Z_2$ orbifold twists.

It is important to emphasise that the example discussed here in detail is a particular example 
that provides insight into the inner working of the spinor-vector duality map. 
However, as examplified in fig. \ref{den} and table \ref{svdcounting},
the SVD applies to the 
wider space of string vacua with $N=2$ and $N=1$ spacetime supersymmetry \cite{svd1, svd2, fkrneq2},
which typically can be of the order of $10^{12}$ distinct models. The analysis in these
papers utilises the free fermionic formulation, which somewhat obscures the role played 
by the geometrical moduli fields. It is demonstrated in \cite{svd2, fkrneq2} 
in terms of the GGSO projection 
coefficients of the one--loop partition function that the SVD always exists in this space of 
vacua. The bosonic analysis in \cite{aft, ffmt} exposes the role
of the moduli fields and shows that the SVD corresponds to an exchange between two Wilson lines. 
Furthermore, as demonstrated in \cite{ffmt} and discussed further below, the SVD can be
interpreted as a map between two Wilson lines, which is induced by the spectral flow operator 
of the $N=2$ worldsheet supersymmetry in the bosonic sector of the heterotic--string. 
The SVD can then be interpreted to arise from the breaking of the $N=2$ worldsheet 
supersymmetry. At the enhanced symmetry point, with unbroken $N=2$ worldsheet supersymmetry, 
the gauge symmetry is enhanced to $E_6$. In this case the spectrum is self--dual under the SVD and 
the spectral flow operator acts as a generator of $E_6$, exchanging between the spinorial and 
vectorial representations in the decomposition of the $E_6$ representations under the 
$SO(10)\times U(1)$ subgroup. This picture then generalises to string vacua
with interacting internal CFTs \cite{afg}, which utilise the Gepner construction 
of such vacua \cite{gepner}. Starting with string vacua with $(2,2)$ worldsheet supersymmetry
and $E_6$ gauge symmetry, the $N=2$ worldsheet supersymmetry on the bosonic side is
broken with a Wilson line and the spectral flow operator then induces the transformation which 
extend the SVD to these cases \cite{afg}. It is important, however, to emphasise that the bosonic
representation that I discussed in detail here is crucial for seeking the imprint of the SVD in 
the effective field theory limit, just as was the case in the case of mirror symmetry. 

The technical details of the relation between the discrete torsion and the Wilson line realisations 
of the SVD are provided in ref. \cite{ffmt}. For our purpose here, it is sufficient to 
realise that in terms of the toroidal background fields, there exist choices 
of the background fields that gives rise to the spectra of the dual models. In those 
terms the $Z_2$ twist action of the internal coordinates is given by eq. (\ref{z2twist}), whereas
the dual Wilson lines are given by 
\begin{align}\label{WL+}
		g = (0,0,0,0,0,1|0,0|1,0,0,0,0,0,0,0).
\end{align}
and 
\begin{align}\label{WL-}
	    g = (0,0,0,0,0,0|1,0|1,0,0,0,0,0,0,0).
\end{align}
Furthermore, the map between the two can be represented in terms of a spectral flow 
operators \cite{ffmt}. A representation of the duality in terms of the spectral flow
operator is discussed further below by using an alternative set of free fermion basis 
vectors. I emphasise, however, that to relate the worldsheet symmetries to the properties
of the effective field theory limit of the string compactifications is facilated by 
using the bosonic data in the form of eqs. (\ref{WL+}) and (\ref{WL-}). The interpretation of the
worldsheet data in the effective field theory limit is often obscured, as is the case, for example,
in the case of mirror symmetry. For this purpose, the representation of the SVD in terms 
of the Wilson lines is particularly useful.

Mirror symmetry was initially observed in worldsheet CFT constructions of string
compactifications, and the profound implications for the geometrical complex manifolds
that are utilised in the effective field theory limit of the string compactifications
was subsequently understood. In particular, mirror symmetry facilitates the counting 
of intersections between sub--surfaces of the complex manifolds, in a field known
as enumerative geometry, which are otherwise notoriously difficult to perform. The 
calculation is facilitated by the relation of the intersection curves to the calculation 
of the Yukawa couplings between the string states. Thus, the worldsheet constructions 
provide a useful tool to study the properties of the string vacua in the effective field theory limit. 

In a similar spirit, it is interesting to explore the implications of spinor--vector duality in 
the effective field theory limit of the string compactifications. 
Similarly, to mirror symmetry, where the modular properties of the worldsheet
string theory have profound implications in the effective field
theory limit on the properties of complex manifolds, the SVD may have similar implications for 
compactifications on complex manifolds with the vector bundles that correspond to the 
gauge degrees of freedom in the worldsheet realisations.
In the case of the $Z_2\times Z_2$ orbifold, 
whereas in the case of mirror symmetry 
the discrete torsion is between two $Z_2$ twists of the internal
compactified torus, in the case of spinor--vector duality the
discrete torsion is between the internal $Z_2$ twist and the $Z_2$
Wilson line that breaks the worldsheet supersymmetry from $(2,2)$ to
$(2,0)$. In the past year we demonstrated the realisation of the
SVD in the effective field theory limit of compactifications to five \cite{5dsvd}, 
and six \cite{6dsvd}, dimensions, by studying resolutions of the orbifold singularities,
whereas the analysis of spinor--vector duality in compactifications to four dimensions
requires further development of the resolutions tools \cite{taming}. 
The SVD therefore
provides a tool to explore the effective field theory limit of $(2,0)$
string compactifications. 
The SVD can serve as a probe of the moduli spaces of heterotic--string compactifications with 
$(2,0)$ worldsheet supersymmetry.
While the moduli spaces of string compactifications
with standard embedding and $(2,2)$ worldsheet supersymmetry
are fairly well understood,
the moduli spaces of $(2,0)$ models are obscured. 
Recently, we demonstrated the viability
of this approach in the case of compactifications to five and six dimensions
\cite{5dsvd,6dsvd}, where the effective field theory limit is obtained by resolving the
orbifold singularities. In this context, the worldsheet description serves as a
guide to guess how the worldsheet description should
be interpreted in the effective field theory limit. It is
noted that the SVD in the worldsheet formalism generalises to string
compactifications
with interacting internal CFTs \cite{afg},
as well as to cases with more discrete torsions \cite{aft}.

Further insight into the structure underlying the SVD is revealed
by breaking the untwisted NS symmetry from $SO(16)\times SO(16)$
to $SO(8)^4$ and generating the $SO(10)$ or $SO(12)$ symmetries
by enhancements \cite{fkrneq2}\footnote{details of the free fermionic 
formalism can be found in the original literature \cite{fff} and are not 
repeated here. In the notation used in eq. (\ref{basis2}) and (\ref{basis}) 
the entries in the curly brackets denote the periodic worldsheet fermions}.
This is obtained by defining
four basis vectors with four non--overlapping sets of four periodic
fermions, as shown in the set of boundary condition
basis vectors given in eq. (\ref{basis2}):
\begin{eqnarray}
v_1=1&=&\{\psi^\mu,\
\chi^{1,\dots,6},y^{1,\dots,6}, \omega^{1,\dots,6}| \nn\\
& & ~~~\bar{y}^{1,\dots,6},\bar{\omega}^{1,\dots,6},
\bar{\eta}^{1,2,3},
\bar{\psi}^{1,\dots,5},\bar{\phi}^{1,\dots,8}\},\nn\\
v_2=S&=&\{\psi^\mu,\chi^{1,\dots,6}\},\nn\\
v_{3}=z_1&=&\{\bar{\phi}^{1,\dots,4}\},\nn\\
v_{4}=z_2&=&\{\bar{\phi}^{5,\dots,8}\},\nn\\
v_{5}=z_3&=&\{\bar{\psi}^{1,\dots,4}\},\nn\\
v_{6}=z_0&=&\{\bar{\eta}^{0,1,2,3}\},\nn\\
v_{7}=b_1&=&\{\chi^{34},\chi^{56},y^{34},y^{56}|\bar{y}^{34},
\bar{y}^{56},\bar{\eta}^0,\bar{\eta}^{1}\},\label{basis2}
\end{eqnarray}
The models generated by the basis (\ref{basis2})
preserve $N=2$ space--time supersymmetry,
as the single supersymmetry breaking vector $b_1$
is included in the basis.

I refer to the basis vectors in eq. (\ref{basis2}) as modular basis vectors.
In this case the $Z_2$ twist of the $b_1$ basis vector 
that acts in the internal dimensions, breaks one $SO(8)$ symmetry to
$SO(4)^2$. The modular basis vector $z_0$,  which is charged under this
$SO(8)$, acts as a spectral flow operator, in a similar way to the $S$--spectral flow
operator that mixes between the spacetime supersymmetric multiplets
on the supersymmetric side of the heterotic-string. The GUT symmetries on the other 
hand are realised in the string models by enhancements. In this case the basis vector
$z_3$, subject to specific choices of GGSO projection coefficients,
enhances the untwisted $SO(8)\times SO(4)$ symmetry to $SO(12)$ (or $SO(8)\times SO(2)\rightarrow SO(10)$ if 
an additional twist basis vector $b_2$ is included). 
In the (2,2) vacua
with enhanced $E_6$ (or $E_7$) symmetry, the $z_0$  basis vector acts
as a symmetry generator of the enhanced symmetry and exchanges between the
$SO(10)\times U(1)$ multiplets inside the $E_6$ representations.
When the $E_6$ symmetry is broken to $SO(10)\times U(1)$, the spectral
flow operator induces the map between the dual Wilson lines and hence
between the two spinor--vector dual vacua \cite{fkrneq2, ffmt}.

The two spectral flow basis vectors, the $S$--vector and the $z_0$--vector, are mere two examples
of a much richer symmetry structure. This much richer symmetry structure
can be glimsped by compactifying the string models to 2 dimensions. In this case the 
target space on the bosonic side of the heterotic--string is 24 dimensional. The $z_{0,1,2,3}$ basis 
vectors in eq. (\ref{basis2}) each contains 4 periodic worldsheet fermions. 
The 24 dimensional lattice in 2 dimensions can therefore be divided into 
6 such non--overlapping basis vectors. The symmetry structure of 24 dimensional lattices
is a topic of much interest. What role this symmetry structure plays in the phenomenological and 
mathematical properties of string theory is yet to be unravelled. An embryonic
attempt to explore this rich symmetry structure in the context of the phenomenological
free fermionic models was discussed in ref. \cite{panos}. Here I give a brief account of these investigations. 

In the light--cone gauge, the worldsheet free fermions of the heterotic--string in two dimensions (in the usual notation \cite{fff,panos}) are:
$\chi^i,y^i, \omega^i, i=1,\dots,8$ (real left-moving fermions)
and
$\bar{y}^i,\bar{\omega}^i, i=1,\dots,8$ (real right-moving fermions),
${\bar\psi}^A, A=1,\dots,4$, 
$\bar{\eta}^B, B=0,1,2,3$, 
$\bar{\phi}^\alpha,
\alpha=1,\ldots,8$ (complex right-moving fermions).
The left-- and right--moving real fermions are 
combined into complex fermions as 
$\rho_i=1/\sqrt{2}(y_i+i\omega_i), ~i=1,\cdots,8$,
${\bar\rho}_i=1/\sqrt{2}({\bar y}_i+i{\bar\omega}_i), ~i=1,\cdots,4$,
${\bar\rho}_i=1/\sqrt{2}({\bar y}_i+i{\bar\omega}_i), ~i=5,\cdots,8$. 

The models of interest are generated by a maximal set $V$ of 7 basis vectors: 
$$
V=\{v_1,v_2,\dots,v_{7}\},
$$
\begin{eqnarray}
v_1=\mathbf1&=&
\{
\chi^{1,\dots,8},y^{1,\dots,8}, \omega^{1,\dots,8}| \nn\\
& & ~~~\bar{y}^{1,\dots,8},\bar{\omega}^{1,\dots,8},
\bar{\eta}^{0,1,2,3},
\bar{\psi}^{1,\dots,4},\bar{\phi}^{1,\dots,8}\},\nn\\
v_2=H_L&=&\{\chi^{1,\dots,8}, y^{1,\dots,8}, \omega^{1,\dots,8}
\},\nn\\
v_{3}=z_1&=&\{\bar{\phi}^{1,\dots,4}\},\nn\\
v_{4}=z_2&=&\{\bar{\phi}^{5,\dots,8}\},\nn\\
v_{5}=z_3&=&\{\bar{\psi}^{1,\dots,4}\},\nn\\
v_{6}=z_4&=&\{\bar{\eta}^{0,1,2,3}\},\nn\\
v_{7}=z_5&=&\{\bar{y}^{1,\dots,4},\bar{\omega}^{1,\dots,4}\},
\label{basis}
\end{eqnarray}
and the corresponding matrix of one--loop Generalised GSO projection coefficients:

\begin{equation}
{\bordermatrix{
        &{\bf 1}&H_L & & z_1  &z_2&z_3 & z_4&z_5 \cr
 {\bf 1}&     -1&  -1& & +1   & +1 & +1  & +1     & +1    \cr
     H_L&     -1&  -1& & \pm1  & \pm1 &\pm1  &  \pm1     &  \pm1      \cr
        &       &    & &       &     &     &         &            \cr
     z_1&     +1&\pm1 & &  +1  &  \pm1 & \pm1  &  \pm1     &  \pm1    \cr
     z_2&     +1&\pm1 & & \pm1 &   +1 & \pm1  &  \pm1     &  \pm1    \cr
     z_3&     +1&\pm1 & & \pm1 &  \pm1 &  +1  &  \pm1     & \pm1    \cr
     z_4&     +1&\pm1 & & \pm1 &  \pm1 & \pm1  &   +1     & \pm1    \cr
     z_5&     +1&\pm1 & & \pm1 & \pm1 &\pm1   &  \pm1     &   +1   \cr}}
\label{phasesmodel1}
\end{equation}

Following the usual free fermion methodology \cite{fff} the range of symmetries
can be classified with varying numbers of basis vectors. Massless bosons are obtained 
from the Neveu--Schwarz sector; the $z_i$ sectors and the $z_i+z_j$, $i=1,\cdots, 6,~
i\ne,j$, where $z_6={\bf 1} +H_L+z_1+z_2+z_3+z_4+z_5= \{{\bar\rho}^{5,6,7,8}\}$. For example, with 
the five basis vectors $\{{\bf 1}, H_L, z_1,z_2,z_3\}$
the configurations of the symmetry groups are summarised in table
\ref{2dconfig}.

\begin{table}[H] 
\caption{Symmetry configurations in compactifications to 2 dimensions.
}
\label{2dconfig}
\setlength{\tabcolsep}{4mm}
%\begin{center}
\begin{tabularx}{\textwidth}{c c c c c c  r}
\toprule
{$\cc{z_1}{H_L}$} & {$\cc{z_2}{H_L}$} & {$\cc{z_3}{H_L}$} &
{$\cc{z_1}{z_2}$} & {$\cc{z_1}{z_3}$} & {$\cc{z_2}{z_3}$} &
\textbf{Gauge Group G}\\\midrule
 +  &  +  &  +  &   +   &   +   &  +  & $SO(24) \times SO(24)$\\
 +  &  +  &  +  &   +   &   +   & $-$ & $SO(8) \times SO(16)\times SO(24) $\\
 +  &  +  & $-$ &   +   &   +   &  +  & $SO(16)\times SO(32)$  \\
$-$ & $-$ &  +  &   +   &   +   &  +  & $SO(8)\times SO(40)$   \\
$-$ & $-$ &  +  &  $-$  &   +   &  +  & $E_8 \times SO(8)\times SO(24)$  \\
$-$ & $-$ & $-$ &   +   &   +   &  +  & $SO(48)$  \\
$-$ & $-$ & $-$ &  $-$  &   +   &  +  & $E_8\times SO(32)$  \\
\bottomrule
\end{tabularx}
\end{table}

%\begin{table}
%\begin{center}
%\begin{tabular}{c c c c c c | r}
%$\cc{z_1}{H_L}$ & $\cc{z_2}{H_L}$ & $\cc{z_3}{H_L}$ &
%$\cc{z_1}{z_2}$ & $\cc{z_1}{z_3}$ & $\cc{z_2}{z_3}$ &
%Gauge group G\\\hline
% +  &  +  &  +  &   +   &   +   &  +  & $SO(24) \times SO(24)$\\
% +  &  +  &  +  &   +   &   +   & $-$ & $SO(8) \times SO(16)\times SO(24) $\\
% +  &  +  & $-$ &   +   &   +   &  +  & $SO(16)\times SO(32)$  \\
%$-$ & $-$ &  +  &   +   &   +   &  +  & $SO(8)\times SO(40)$   \\
%$-$ & $-$ &  +  &  $-$  &   +   &  +  & $E_8 \times SO(8)\times SO(24)$  \\
%$-$ & $-$ & $-$ &   +   &   +   &  +  & $SO(48)$  \\
%$-$ & $-$ & $-$ &  $-$  &   +   &  +  & $E_8\times SO(32)$  \\
%\end{tabular}
%\caption{The configuration of the symmetry group with five basis vectors.}
%\label{2dconfig}
%\end{center}
%\end{table}
The untwisted symmetry is
$SO(8)_1\times SO(8)_2\times SO(8)_3 \times SO(24)$.
There are a total of six independent phases in this case 
producing 64 distinct possibilities, which give rise to
seven distinct configurations shown in the table above. 
The symmetry structure of the worldsheet string constructions may have profound
phenomenological implications. It is aparent that we have thus far only glimpsed some 
of these potential implications. For example, we can consider modular maps, akin to the 
supersymmetry map $S$ and the SVD map $z_0$ that can induce maps between supersymmetric 
and non--supersymmetric vacuum configurations \cite{nonsusy}. 
This map is illustrated in the table shown in eq. (\ref{undernahe})
\beqn
 &&\begin{tabular}{c|c|ccc|c|ccc|c}
 ~ & $\psi^\mu$ & $\chi^{12}$ & $\chi^{34}$ & $\chi^{56}$ &
        $\bar{\psi}^{1,...,5} $ &
        $\bar{\eta}^1 $&
        $\bar{\eta}^2 $&
        $\bar{\eta}^3 $&
        $\bar{\phi}^{1,...,8} $ \\
\hline
\hline
      {\bf~1} & ~1 &~1&1&1 &~1,...,1 &~1 &~1 &~1 &~1~1~1~1~1~1~1~1 \\
          \ds & ~1 &~1&1&1 &~0,...,0 &~0 &~0 &~0 &~0~0~1~1~1~1~0~0 \\
\hline
  ${b}_1$ & ~1 &~1&0&0 &~1,...,1 &~1 &~0 &~0 &~0~0~0~0~0~0~0~0 \\
  ${b}_2$ & ~1 &~0&1&0 &~1,...,1 &~0 &~1 &~0 &~0~0~0~0~0~0~0~0 \\
  ${b}_3$ & ~1 &~0&0&1 &~1,...,1 &~0 &~0 &~1 &~0~0~0~0~0~0~0~0 \\
\end{tabular}
   \nonumber\\
   ~  &&  ~ \nonumber\\
   ~  &&  ~ \nonumber\\
     &&\begin{tabular}{c|cc|cc|cc}
 ~&      $y^{3,...,6}$  &
        $\bar{y}^{3,...,6}$  &
        $y^{1,2},\omega^{5,6}$  &
        $\bar{y}^{1,2},\bar{\omega}^{5,6}$  &
        $\omega^{1,...,4}$  &
        $\bar{\omega}^{1,...,4}$   \\
\hline
\hline
    {\bf~1} &~1,...,1 &~1,...,1 &~1,...,1 &~1,...,1 &~1,...,1 &~1,...,1 \\
    \ds     &~0,...,0 &~0,...,0 &~0,...,0 &~0,...,0 &~0,...,0 &~0,...,0 \\
\hline
${b}_1$ &~1,...,1 &~1,...,1 &~0,...,0 &~0,...,0 &~0,...,0 &~0,...,0 \\
${b}_2$ &~0,...,0 &~0,...,0 &~1,...,1 &~1,...,1 &~0,...,0 &~0,...,0 \\
${b}_3$ &~0,...,0 &~0,...,0 &~0,...,0 &~0,...,0 &~1,...,1 &~1,...,1 \\
\end{tabular}
\label{undernahe}
\eeqn
The map introduced in the model shown in eq. (\ref{undernahe}) is obtained 
by augmenting the SUSY generator basis vector $S$ by four periodic hidden sector
worldsheet fermions. It refered to as as the ${\tilde S}$--map. Up to the \ds--map the 
basis vectors in eq. (\ref{undernahe}) are identical to the NAHE--set basis vectors 
\cite{nahe}, and is refered to as the \unahe--set. The effect of the \ds--map
is to make the gravitino massive. The four dimensional models obtained by the 
\ds--map correspond to compactifications of a tachyonic ten dimensional vacuum,
and the tachyonic states are projected out in the four dimensional models \cite{spwsp}. %
Every model constructed with the NAHE--set can be mapped to a \unahe--based model by
the \ds--map, with the model of ref. \cite{stable} providing a concrete example.
The important point to realise is that understanding the string dynamics in the 
early universe mandates the understanding of the string vacua not only in the 
relatively safe and stable limit of the supersymmetric configurations, but rather 
also those of the non--supersymmetric and unstable configurations. 
Such modular maps may be particularly relevant
for developing an understanding of the unstable configurations and the early dynamics of string theory. 

We noted the rich symmetry structure underlying the worldsheet string constructions. 
What are the implications for the effective field theory target space limits? 
%
%
%
%
%In a bosonic representation of the SVD \cite{aft, ffmt} the
%map bwteen the dual vacua results from an exchange of a discrete
%torsion. In this case the analysis shows that for one choice of the
%discrete torsion, the zero modes of the untwisted torus in the
%$N=2$ twisted sector are attached to the spinorial characters of the
%GUT group, whereas for the other choice they are attached to the
%vectorial character. The choice of discrete torsion can be
%represented as the choice of the Wilson line background. 
%In the vacua that possess $N=2$ spacetime supersymmetry
%the transformation between the dual Wilson lines is
%continuous \cite{ffmt}, whereas in those that possess
%$N=1$ it is discrete. The reason that in the $N=1$ case
%the moduli that allow the continuous interpolation in the
%$N=2$ case are projected out. 
%
Turning back to the case of spinor--vector duality, 
we note that in the exact string theory solutions the entire
space of compactifications is connected at the $N=4$ level. The moduli space
is the moduli space of the underlying $N=4$ toroidal compactification.
The observation made in ref. \cite{fknr} is
that the number of chiral models is predetermined by the moduli space of 
the $N=4$ toroidal compactification, {\it i.e.} it is determined by GGSO phases that correspond 
to $N=4$ moduli. That is, the information of the chiral spectrum is predetermined
by the moduli data of the background toroidal lattice.
At the $N=2$, or $N=1$, level the vacua are related by continuous interpolations, 
or discrete mappings, of the Wilson line moduli. This is because in the $N=1$ case
the interpolating moduli are projected out from the spectrum and the mapping 
between the dual models can only be discrete. The important observation, however, 
is that the space of models is determined by the moduli space of the underlying 
$N=4$ compactifications. The additional $Z_2\times Z_2$ projections produce different
spectra depending on the point in the $N=4$ moduli space where the projection 
is made. This is crucial from the point of view of the effective field theory limit. 
From the worldsheet compactification, we may conjecture that every string vacua 
should be connected either by a continuous interpolation or discrete mapping 
by moduli of the underlying toroidal compactifications. For example, it was noted above
that the spinor-vector duality mapping is induced by the modular map, which is induced
by the basis vector $z_0$ that preserves the $N=4$ spacetime time supersymmetry. 
This is similar to the case of the supersymmetry generator in the $N=1$ and $N=2$
theories that are generated by the basis vector $S$, which itself preserves $N=4$
spacetime supersymmetry. We may speculate that the entire space of $(2,0)$ 
compactifications is connected by interpolations or by discrete transformations, 
where the cases of discrete transformations corresponds to the cases where the 
$N=4$ moduli are projected out. This view is supported by the well known observations 
that the known ten dimensional vacua are connected via orbifolds or by interpolations
in lower dimensions \cite{ginspargvafa, itoyamataylor, ginsparg}. 
%The spinor--vector duality which is observed in $Z_2\times Z_2$ orbifold
%compactifications generalises to exact string solutions with interacting
%internal CFT \cite{afg}.

%As discussed above the spinor vector duality is induced by a modular
%basis vector, and reflects the modular properties of the partition
%function. The maximal symmetric space of this kind is obtained in
%compactifications to two dimensions, where the MSDS symmetry of
%cite{msds} is an example of such a symmetry that results
%from theta function identities in compactifications to two
%dimensions. The underlying reason is the large symmetry struture of 24
%dimensional lattices and it is an interesting prospect that the
%spinor--vector duality may similarly be rooted in the rich
%symmetry structure of 24 dimensional lattices \cite{panos}.

The question is what can we learn from these symmetries of the worldsheet compactifications
about the effective field theory moduli spaces. The imprint of these symmetries in the
form of mirror symmetry and spinor--vector duality has already been noted. But the connectedness of the worldsheet space of vacua is yet to leave its mark. 
At this stage we can only speculate what this may entail. For example, we may question whether every Calabi--Yau manifold has its mirror, and whether every $(2,0)$ Calabi--Yau manifold
should exhibit the spinor--vector duality. An effective field theory limit that does not exhibit these properties is necessarily in the swampland. The question is basically
how do the symmetries of the worldsheet description constrain their effective field
theory limits. In this respect, the SVD and mirror symmetry provide a top-down
approach to the question posed by the swampland program, {\it i.e.} what is the 
relation between exact string solutions and their effective field theory limits. 
Whereas the swampland program aims to explore when does an effective field 
theory description of quantum gravity have an embedding in string theory, 
the duality program seeks to explore how the symmetries of the exact string solutions 
constrain the effective field theory limit of string vacua. As we have seen the 
space of symmetries of the worldsheet compactifications is vast, relating not only
supersymmetric vacua, but also supersymmetric and non--supersymmetric vacua via
the \ds--map. So the road ahead is treacherous and one should not expect
a short journey. Nevertheless, on the way there are many interesting concrete and 
interesting questions to explore, {\it e.g.} the moduli spaces of (2,0) string 
compactifications; the calculation of correlators on dual manifolds; the interpretation of the parameters of the worldsheet theories in the effective field theory limit. The ultimate questions is that of completeness. What is the complete set of symmetries of the 
worldsheet theories and do they completely constrain the effective field theory limit of quantum gravity? 

\section{String dualities and fundamental principles underlying quantum gravity}

The vast space of symmetries underlying the worldsheet string 
vacua was noted in the previous section. Shedding light on this vast space
provides ample number of questions to explore and investigate. Another approach
is to try to extract from the string symmetries, fundamental principles that may
underlie quantum gravity and use them as a starting point to try to formulate the 
theory. $T$--duality is thought to be a basic property of string theory that arises 
due to its extended nature. 
We have seen that  
spinor--vector duality can be thought of as an extention of $T$--duality to include 
symmetries under the mappings of Wilson line moduli rather than the internal moduli 
of the torus. We may continue to explore $T$--duality as a basic symmetry,
{\it e.g.} the double geometry formalism of ref. \cite{hull}. 
Another possible interpretation of $T$--duality is as a form of phase--space duality in compact space, where the duality is exchanging position modes (winding modes) with 
momentum modes. We can then promote phase--space duality to a level of a fundamental 
principle and explore the consequences. 

This is in essence the program that was pursued in the formulation of quantum 
mechanics from an equivalence postulate \cite{fm2, epoqm,bfm}. The starting point for the development
of this approach was the requirement of manifest phase space duality
that arises due to the involutive nature
of Legendre transformations \cite{fm2}. I focus here on the one dimensional stationary case. 
The basic relation
between the phase space variables is defined via the generating
function $S_0$ and its Legendre dual $T_0$
\beq
p = \frac{\partial S_0}{\partial q} 
~~~\hbox{and} ~~~
q = \frac{\partial T_0}{\partial p} 
\label{peqdsdq}
\eeq

Setting the condition ${\tilde S}_0({\tilde q})=
S_0(q)$, {\it i.e.} that $S_0$ transforms as a 
scalar function under the transformations $q\rightarrow {\tilde q(q)}$, 
we obtain
manifest {$p\leftrightarrow q$} -- 
              {$S_0\leftrightarrow T_0$} duality with \cite{fm2,epoqm}
\begin{eqnarray}
~~~~~p~&=&~{{\partial S_0}\over{\partial q}}~~~~~~~~~~~~~~~~~~~~~~~~~~~
~~~~~~~~~~q~=~{{\partial T_0}\over{\partial p}}\nonumber\\
~~~~~S_0~&=&~p{{\partial T_0\over\partial p}}- T_0~~~~~~~~~~~~~~~~~~~~~~~~~~~
T_0~=~q{{\partial S_0\over\partial q}}- S_0\nonumber
\end{eqnarray}
\begin{equation}
\left({{\partial^2~~} \over\partial S_0^2}+{U(S_0)}\right)
\left({{q\sqrt p}\atop \sqrt{p}}\right)=0
~~~~~~~~~~
\left({{\partial^2~~}\over\partial T_0^2}+{\cal V}(T_0)\right)
\left({{p\sqrt q}\atop \sqrt{q}}\right)=0\label{diffs0eq}
\end{equation} 
where a dual second order differential equations in the third 
line are associated with the dual Legendre transformations
in the second line. The potential function $U(S_0)$ is given by
$U(S_0)= \{q, S_0\}/2$, where $\{h(x),x\}=h^{\prime\prime\prime}/h^\prime - (3/2)
(h^{\prime\prime}/h^\prime)^2$ denotes the Schwarzian derivative. 

There are several important points to note. 
The first is the 
the existence of self--dual states, which are 
simultaneous solutions of the dual pictures. 
For these states we have 
$S_0=-T_0+constant$, and 
\beq
S_0(q)=\gamma\ln\gamma_qq~~~~~~~~~~~~~~~~~T_0(p)=\gamma\ln\gamma_p p
\label{selfdualstates}
\eeq
Here, $S_0~+~T_0~=~pq~=~\gamma$, with
$\gamma_q\gamma_p\gamma={\rm e}$ and $\gamma_q$, $\gamma_p$ are constants. 
As the Legendre transformation is undefined for linear functions,
the case of physical states with $S_0=Aq+B$ with constants $A$ and $B$
is excluded. As we will see below, these cases coincide precisely with 
the self--dual states, and quantum mechanics rectifies this inconsistency of the 
classical limit. For now we note the self--dual solutions that are given in 
eq. (\ref{selfdualstates}). 

From the basic properties of the 
Schwarzian derivative, it is further noted that the potential function $U(S_0)$ is invariant  under the 
$GL(2,C)$--transformations
\beq
{\tilde q}={{A q + B}\over{ C q + D}}~,~~~
{\tilde p}= \rho^{-1}(C q + D)^2 p, 
\label{gl2ctrans}
\eeq
where $\rho=A D - B C\ne0$. However, under arbitrary coordinate transformations 
$q\rightarrow {\tilde q}= v(q)$ we have ${\tilde U}({\tilde s})\ne U(s)$, 
whereas the condition ${\tilde S}({\tilde q})= S_0(q)$ implies that the differential
equation eq. (\ref{diffs0eq}) is covariant under the coordinate transformations. This suggests that different
physical systems labelled by different potentials can be connected by
coordinate transformations. This suggests the fundamental equivalence postulate 
\cite{fm2, epoqm}: 

{\it Given two physical systems labelled by
potential functions $W^a(q^a)\in H$ and 
$W^b(q^b)\in H$, where $H$ denotes the space
of all possible $W$'s, there always exists 
a coordinate transformations  $q^a\rightarrow q^b=v(q^a)$
such that $W^a(q^a)\rightarrow W^{av}(q^b)=W^b(q^b)$. }

It follows that there should always exist 
a coordinate transformation connecting any state
to the trivial state $W^0(q^0)=0$. Conversely,
any nontrivial state $W\in H$ can be obtained from the 
states $W^0(q^0)$ by a coordinate transformation. 

A natural setting to develop this approach is provided by the classical 
Hamilton--Jacobi formalism. I focus here on the stationary case. 
In the HJ formalism
the physical problem is solved by using canonical 
transformations that map a non--trivial Hamiltonian, 
with nonvanishing kinetic and potential energies, to a trivial Hamiltonian.
The Classical Stationary Hamilton--Jacobi Equation (CSHJE) 
\beq
{1\over{2m}}\left({\partial_q S}_0\right)^2 + W(q)=0, 
\label{cshje}
\eeq
where $W(q)=V(q)-E$, 
provides the solution
and the functional relation between the 
phase space variables is extracted by the relation 
$p=\partial_q S_0$, where $S_0$ is the solution of the 
CSHJE equation. 
We can pose a similar question, but imposing 
the functional relations $p=\partial_q S_0(q)$ on the trivialising
transformation $q\rightarrow q^0(q)$ and $S_0^0(q^0)=S_0(q)$.
The CSHJE is not consistent with this 
procedure because 
the free state $W^0(q^0)\equiv0$ is a fixed point under the coordinate 
transformations \cite{fm2,epoqm}. It is noted that this state correspond 
to the self--dual state under phase space duality. Consistency of the 
equivalence postulate therefore implies that the CSHJE has to be 
be modified. Focusing on the stationary case, the most general 
modification is given by
\beq
{1\over{2m}}\left({\partial_q S}_0\right)^2 + W(q) + Q(q) =0. 
\label{qshje}
\eeq
By the equivalence postulate eq. (\ref{qshje})
is covariant under general coordinate transformations. 
This is achieved if the combination $(W+Q)$ 
transforms as a quadratic differential under general coordinate 
transformations. Furthermore, 
all nontrivial states should be obtained from the state
$W^0(q^0)$ by a coordinate transformation. 
It follows that each of the functions 
$W(q)$ and $Q(q)$ transform
as quadratic differentials up to an additive term, {\it i.e.}
under $q\rightarrow {\tilde q}(q)$ we have, 
\beqn
W(q)\rightarrow  {\tilde W}({\tilde q}) & = &
              \left( {{\partial q}\over {\partial{\tilde q}}}\right)^2
                        {W}({q}) + (q;{\tilde q})\nonumber\\
Q(q)\rightarrow  {\tilde Q}({\tilde q}) & = &
              \left( {{\partial q}\over {\partial{\tilde q}}}\right)^2
                        {Q}({q}) - (q;{\tilde q}).\nonumber
\eeqn
and 
\beq
(W(q)+Q(q))\rightarrow  ({\tilde W}({\tilde q})+
                         {\tilde Q}({\tilde q}))= 
              \left( {{\partial q}\over {\partial{\tilde q}}}\right)^2
                         (W(q)+Q(q))\nonumber
\eeq 
It is seen that all non--trivial potential functions $W^a(q^a)$ can 
be generated from the trivial state $W^0(q^0)\equiv0$ by 
coordinate transformations $q^0\rightarrow q^a$, with 
$W^a(q^a)\equiv(q^0;q^a)$, {\it i.e.} they arise from the
inhomogeneous term. 
Considering the transformation $q^a\rightarrow q^b\rightarrow q^c$
versus $q^a\rightarrow q^c$ gives rise to the cocycle
condition on the inhomogeneous term 
\beq
(q^a;q^c)= \left({{\partial q^b}\over {\partial q^c}}\right)^2
           \left[ (q^a;q^b) - (q^c;q^b)\right]. 
\label{cocycle}
\eeq
The cocycle condition, eq. (\ref{cocycle}), underlies the equivalence
postulate, and embodies its underlying symmetries. In particular, it
is invariant under the M\"obius transformations, 
\beq
(\gamma(q^a);q^b)=(q^a;q^b),
\label{am21}\eeq
where
\beq
~~~~~~~~~~~~~~~~~~~~~
\gamma(q)={{Aq+B}\over {Cq+D}}~~~~~\hbox{and}~~~~~
\left(\begin{array}{c}A\\C\end{array}\begin{array}{cc}B\\D\end{array}
\right)\in GL(2,{C}).
\label{mobiustrans}
\eeq
The intimate connection between the Equivalence Postulate of Quantum Mechanics and
the phase space duality ought to be emphasised.
This is elucidated by considering further the 
structure of the formalism in the one dimensional
case \cite{fm2}. The one dimensional case captures the symmetry structures that 
underlie the formalism and is amenable, rather straightforwardly,
for generalisations to higher dimensions in Euclidean and 
Minkowski spacetimes \cite{bfm}. Preserving these symmetry structures is 
the key to the generalisations, but the basic physical 
features can already be gleaned from the one dimensional 
structure. A basic identity in the one--dimensional stationary
case takes the form of a difference between two Schwarzian 
derivatives 
\beq
\left(\frac{\partial S(q)}{\partial q}\right)^2=
{\beta^2\over 2}
\left(\left\{{\rm e}^{{{2i}\over\beta}{S}},q\right\}-
\left\{S,q\right\}\right)
\label{schwarzianidentity}
\eeq
where
$\{f,q\} =f^{\prime\prime\prime}/f^\prime-3(f^{\prime\prime}/f^\prime)^2/2$ denotes
the Schwarzian derivative and $\beta$ is a constant
with the dimension of an action. Making the identification
\beq
W(q)= V(q)-E = 
-{\beta^2\over {4m}}\left\{{\rm e}^{{{(i2S_0)}\over \beta}},q\right\},
\label{wqeqvqminuse}
\eeq
and 
\beq
Q(q) = {\beta^2\over {4m}}\left\{S_0,q\right\}, 
\label{qq}
\eeq
we have that $S_0$ is the solution of the Quantum Stationary Hamilton--Jacobi
equation (QSHJE), 
\beq
{1\over {2m}}\left({{\partial S_0}\over 
{\partial q}}\right)^2 + V(q) - E + {\hbar^2\over{4m}}
\left\{S_0,q\right\} = 0 .
\label{qshje2}
\eeq
From, eq. (\ref{wqeqvqminuse}), and
the properties of the Schwarzian derivative, it follows that $S_0$, 
the solution of the QSHJE eq. (\ref{qshje2}) is given by 
(see also \cite{floyd}),
\beq
{\rm e}^{{2i\over \beta}S_0}= \gamma(w)= 
{{Aw+B}\over {Cw +D}} = {\rm e}^{i\alpha}
{{w+i{\bar\ell}}\over {w-i\ell}}
\label{ei2s0overbeta}
\eeq
where $\ell= \ell_1+i\ell_2$; $\{\alpha, \ell_1, \ell_2\}\in R$.  
Here $w=\psi^D/\psi$ and $\psi^D$ and $\psi$ are two 
linearly independent solutions
of a second order differential equation given by
\beq
\left(-{\beta^2\over {2m}} {\partial^2\over 
{\partial q}^2} + V(q) - E \right)\psi(q)=0
\label{se}
\eeq
{\it i.e.} $\psi^D$ and $\psi$ are the two solutions of 
the Schr\"odinger equation
and we can identify $\beta\equiv\hbar$.
Eq. (\ref{schwarzianidentity}) follows from the requirement
that the QHJE is covariant under coordinate transformations
\cite{fm2}.

The self--dual solutions under phase--space duality coincide with
the $W^0(q^0)\equiv 0$ states of the Quantum Hamilton--Jacobi Equation (QHJE). 
The important observation is that the consistency of phase--space duality, 
as well as the EPOQM, requires a departure from classical mechanics that 
mandates that $S_0\ne \gamma(q)$, where $\gamma(q)$ is a M\"obious 
transformation of $q$ and in turn that the quantum potential 
$Q(q)=(\beta^2/4m) \{S_0, q\}$ is never vanishing. It is further noted
that the solutions of eq. (\ref{qshje2}) precisely coincide with the
self--dual solutions in eq. (\ref{selfdualstates}). 

The key features of the formalism are encoded in the Schwarzian
identity (\ref{schwarzianidentity}) and cocycle condition
(\ref{cocycle}). These two key elements generalise to 
any number of dimensions with Euclidean or Minkowski metrics \cite{bfm}. 
The Schwarzian identity generalises to a quadratic identity 
given by 
\beq 
\alpha^2(\nabla S_0)^2=
{\Delta(R{\rm e}^{\alpha S_0})\over 
R{\rm e}^{\alpha S_0}}-{\Delta R\over R}-
{\alpha\over R^2}\nabla\cdot(R^2\nabla S_0), 
\label{ddidentity}
\eeq
which holds for any constant $\alpha$ and any functions $R$ and $S_0$. 
Then, if $R$ 
satisfies the continuity equation 
\beq
\nabla\cdot(R^2\nabla S_0)=0, 
\label{conteq}
\eeq
and setting $\alpha=i/\hbar$, we have 
\beq
{1\over2m}(\nabla S_0)^2=-{\hbar^2\over2m}{\Delta(R{\rm e}^{{i\over 
\hbar} S_0})\over R{\rm e}^{{i\over\hbar}S_0}}+
{\hbar^2\over2m}{\Delta R\over R}. 
\label{identity2}
\eeq 
In complete analogy with the one dimensional case we make identifications, 
\beqn
W(q)=V(q)-E& = &{\hbar^2\over2m}{\Delta(R{\rm e}^{{i\over\hbar}S_0})\over 
R{\rm e}^{{i\over \hbar}S_0}}, 
\label{ddwq}\\
Q(q)& =& -{\hbar^2\over2m}{\Delta R\over R}. 
\label{identity3}
\eeqn
Eq. (\ref{ddwq}) implies the $D$--dimensional Schr\"odinger equation
\beq
\left[-{\hbar^2\over2m}{\Delta}+V(q)\right]\Psi=E\Psi. 
\label{ddschroedingereq}
\eeq
and the general solution
\beq
\Psi= R(q) \left( A {\rm e}^{{i\over \hbar} S_0} + 
B {\rm e}^{-{i\over \hbar} S_0}\right), 
\label{ddwavefunction}
\eeq
and similarly in the relativistic case the Schwarzian
identity generalises to 
\beq
\alpha^2(\partial S)^2={\Box(R{\rm e}^{\alpha S})\over R{\rm e}^{\alpha S}} 
-{\Box R\over R}-{\alpha\over R^2}\partial \cdot (R^2\partial S), 
\label{qidentityrel}
\eeq
which holds for any constant $\alpha$ and any functions $R,$ and $S$. 
Then, if $R$ satisfies 
the continuity equation $\partial(R^2\cdot\partial S)=0$, and setting 
$\alpha=i/\hbar$ we have 
\beq 
%{1\over2m}
(\partial S)^2=-{\hbar^2}{\Box(R{\rm e}^{{i\over 
\hbar} S})\over R{\rm e}^{{i\over\hbar} S}}+{\hbar^2}{\Box R\over R}. 
\label{identity2rel}
\eeq
Setting 
\beqn
W(q) ~=~ {mc^2} & = & -{\hbar^2}{\Box(R{\rm e}^{{i\over 
\hbar} S})\over R{\rm e}^{{i\over\hbar} S}} \label{wqrelativistic}\\ 
Q(q) & =& {\hbar^2}{\Box R\over R} \label{qqrelativistic}
\eeqn
reproduces the relativistic Klein--Gordon equation
\beq
\left( {\hbar^2}{\Box}+ mc^2\right) \Psi(q) = 0 \label{rkge}
\eeq
with the general solution
\beq
\Psi  = R(q) ( A  {\rm e}^{{i\over  \hbar} S} + 
               B  {\rm e}^{-{i\over  \hbar} S}). 
\label{rwf}
\eeq
The cocycle condition eq. (\ref{cocycle}) similarly generalises to any
number of dimensions with Euclidean or Minkowski metrics. For example,
in Minkowski spacetime 
setting $q\equiv (ct, q_1, \ldots, q_{D-1})$, 
with $q^v=v(q)$ a general 
transformation of the coordinates, we have  
\beq
(p^v|p)={\eta^{\mu\rho}p_\mu^vp_\rho^v\over\eta^{\mu\rho}p_\mu p_\nu}=
{p^tJ\eta J^tp\over 
p^t\eta p}, 
\label{pvprelativistic}
\eeq
and $J$ is the Jacobian matrix 
\beq
{J^\mu}_\rho={\partial q^\mu\over\partial{q^v}^\rho}. 
\label{relativisticjacobian}
\eeq
Furthermore, we obtain the cocycle condition 
\beq
(q^a;q^c)=(p^c|p^b)\left[(q^a;q^b)-(q^c;q^b)\right], 
\label{relcocycle}
\eeq
and is invariant under $D$--dimensional M\"obius transformations with respect to
Minkowski metric. 
The cocycle condition eq. (\ref{cocycle}) similarly generalises to any number
of dimensions with Euclidean metric and is given by
\beq
(q^a;q^c)=(p^c|p^b)\left[(q^a;q^b)-(q^c;q^b)\right],
\label{cocycleinEspace}
\eeq
where
\beq
(p^v|p)={{\sum_k (p_k^v)^2}\over{\sum_kp_k^2}}={{p^tJ^tJp}\over {p^tp}},
\label{pvp}
\eeq
and 
\beq
J_{ki}={{\partial q_i}\over {\partial q_j^v}},
\label{jacobian}
\eeq
is the Jacobian of the $D$--dimensional 
transformation, $q\rightarrow q^v = v(q)$,
with $S_0^v(q^v)=S_0(q)$ and 
$p_k= {\partial_q S_0}.$
It is shown \cite{bfm} that the cocycle condition eq. 
(\ref{cocycleinEspace}) is invariant under $D$--dimensional 
M\"obius transformations with Euclidean or Minkowski metrics. 

It is therefore seen that the key ingredients of the 
EPOQM formalism generalise to any number of dimensions. 
Furthermore, and crucially, the key symmetry property of 
quantum mechanics in this approach is the invariance
of the cocycle condition under $D$--dimensional 
M\"obius transformations. This is crucial because consistent
implementation of the M\"obius symmetry necessitates that 
spatial space is compact. The reason being that M\"obius 
transformations include a symmetry under reflection with respect 
to the unit sphere \cite{bfm}. This is the fundamental property
of quantum mechanics in the EPOQM formulation. This feature of
the formalism coincide with the property that the quantum 
potential is never vanishing. 

The same basic structure generalises to curved space as well. 
The basic quadratic identity can be written in curved space
in the form. 
\beq
\alpha^2(\partial_\mu S)(\partial^\mu S)=
\frac{
\frac{1}{\sqrt{g}}
\partial_\mu
{\sqrt{g}}
\partial^\mu\left(R{\rm e}^{\alpha S}\right)}
{R{\rm e}^{\alpha S}
}- 
\frac{
\frac{1}{\sqrt{g}}
\partial_\mu
\left(
{\sqrt{g}}
\partial^\mu R\right)}
{R}
-{\alpha\over R^2}
\frac{1}{\sqrt{g}}
\partial_\mu \left( \sqrt{g} R^2\partial^\mu S\right), 
\label{qidincurspa}
\eeq
where $S$ and $R$ are scalar functions. 
We can similarly extend this basic structure in the case of 
fields. In particular, treating the spacetime metric as a field 
we can utilise the Wheeler--deWitt to write a corresponding
QHJE for the spatial part of the metric \cite{holland, gravcon}. This
equation is given by
\beq
\alpha^2
G_{ijkl} 
\frac{\delta S}{\delta g_{ij}}
\frac{\delta S}{\delta g_{kl}}
= 
\frac{1}{R{\rm e}^{\alpha S}}
 G_{ijkl}
\frac{ \delta^2 \left(R{\rm e}^{\alpha S}\right)}
{\delta g_{ij} \delta g_{kl}}
- 
G_{ijkl}
\frac{1}{R}
\frac{  \delta^2 \left(R\right)}
{\delta g_{ij} \delta g_{kl}}
-
\frac{\alpha}{R^2} G_{ijkl} 
\frac{\delta}{\delta g_{ij}}
\left(R^2 \frac{\delta S}{\delta g_{ij}}\right)
\label{wdwid}
\eeq
Following the EPOQM structure the WDW equation corresponding to 
eq. (\ref{wdwid}) is obtained by identifying the 
first part on the right--hand side of eq. (\ref{wdwid})
with the classical potential {\it i.e.} 
\beq
\frac{1}{R{\rm e}^{\alpha S}}
 G_{ijkl}
\frac{ \delta^2 \left(R{\rm e}^{\alpha S}\right)}
{\delta g_{ij} \delta g_{kl}}
= -\sqrt{g}~ ^{3}{\cal R}
\label{wdweq}
\eeq
In eqs. (\ref{wdwid},\ref{wdweq}) $g=\hbox{det} g_{ij}$, 
${^3{\cal R}}$ is the spatial intrinsic curvature, and
$G_{ijkl}=\frac{1}{2}g^{-1/2}(g_{ik}g_{jl}+g_{il}g_{jk}-g_{ij}g_{kl})$
is the supermetric, and
$\Psi\left[g_{ij}({\bf x})\right]=R{\rm e}^{\alpha S}$
is a wavefunctional in the superspace, the ``wavefunction of the 
universe''. The quantum version of the corresponding Hamilton--Jacobi
equation is then given by
\beq
G_{ijkl} 
\frac{\delta S}{\delta g_{ij}}
\frac{\delta S}{\delta g_{kl}}
-
\sqrt{g}~ ^{3}{\cal R} 
- 
G_{ijkl}
\frac{\hbar^2}{R}
\frac{  \delta^2 \left(R\right)}
{\delta g_{ij} \delta g_{kl}}
= 0. 
\label{wdwhje}
\eeq

We next turn to examine these consequences in the context of the
EPOQM. The pivotal property of the EPOQM 
is the M\"obious symmetry that underlies quantum mechanics
in this formalism. The M\"obius symmetry indicates that the 
spatial space is compact. The time coordinate cannot of course 
be compact and it is well known that the M\"obius symmetry 
does not imply compactness in Minkowski space. However, the 
M\"obius symmetry in the EPOQM underlies quantum mechanics
in the nonrelativistic limit, and can only be applied 
consistently if spatial space is compact. On the other hand, 
observations dictate that space is locally flat. These two properties
are compatible with observations if spatial space is, for example, 
$T^3$. On the other hand, as elaborated above the EPOQM dictates 
that $Q(R)\ne 0$. The EPOQM formalism therefore predicts the 
compactness of space as well as the non--vanishing ground
state energy associated with the quantum potential $Q(R)$.
Both properties arise as a basic consequence of the  
M\"obius symmetry that underlies quantum mechanics in the 
EPOQM approach. 
Having demonstrated the robustness of the generalisation of
the Schwarzian identity, eq. (\ref{schwarzianidentity}), 
to any number of dimensions, 
as well as in curved space, we can take a leap of 
faith and propose that the spatial identity in 
eq. (\ref{wdwid}) generalises to general space 
in the form
\beqn
& &
\alpha^2
G_{\mu\nu\eta\rho} 
\frac{\delta S}{\delta g_{\mu\nu}}
\frac{\delta S}{\delta g_{\eta\rho}}
=  \nonumber\\
& & ~~~~~
\frac{1}{R{\rm e}^{\alpha S}}
 G_{\mu\nu\eta\rho}
\frac{ \delta^2 \left(R{\rm e}^{\alpha S}\right)}
{\delta g_{\mu\nu} \delta g_{\eta\rho}}
- 
G_{\mu\nu\eta\rho}
\frac{1}{R}
\frac{  \delta^2 \left(R\right)}
{\delta g_{\mu\nu} \delta g_{\eta\rho}}
-
\frac{\alpha}{R^2} G_{\mu\nu\eta\rho} 
\frac{\delta}{\delta g_{\mu\nu}}
\left(R^2 \frac{\delta S}{\delta g_{\eta\rho}}\right),
\label{wdwlof}
\eeqn
and that in this form this identity can serve as a starting point
for a covariant approach to quantum gravity. Note that in this form
we have not assigned any interpretation to the terms arising in this
identity, which would require further explorations into its 
properties and implications. 

%%CONTINUE FROM HERE

%%%%%%%%%%%%%%%%%%%%%%%%%%%%%%%%%%%%%%%%%%
\section{Conclusions}

The synthesis of gravity with quantum mechanics is likely to occupy theoretical physicists well into the third millennium. It is naive to expect a quick fix. Twentieth century physicists have made deep inroads in our understanding of the material world in the very small and in the very large. More importantly, these inroads are supported by substantial observational data, which is the key to their respective success. Yet, they are not satisfactory. There is a fundamental dichotomy between the mathematical model that describe the observational data in the sub--atomic world versus the mathematical model that describe the observational data in the celestial, 
galactical and cosmological spheres. This dichotomy is particularly glaring in regard to the vacuum. While
gravitationally based observations show that the vacuum energy is highly suppressed, the quantum field theory
models that are used to account for the sub--atomic data predict a vacuum energy which is many orders of magnitude larger.

The available experimental data indicates that the Standard Model provides
viable parametrization of all observational data up to the GUT and Planck scales. The multiplet structure of the Standard Model fermions; the logarithmic evolution of the Standard Model parameters; the longevity 
of the proton and the suppression of the left--handed neutrino masses
strongly support the embedding of the Standard Model states in 
multiplets of 
a Grand Unified Theory, which is realised at the GUT or Planck scales. 
This is the minimal hypothesis that one may infer from the currently 
available observational data. 
However, embedding the Standard Model in a Grand Unified Theory
still leaves too many unexplained parameters. In particular in
the flavour sector of the Standard Model. The 
fundamental origin of these parameters, in particular in
the favour sector, can only be
revealed by unifying the Standard Model with gravity. String theory
is a mundane extension of the quantum field theory framework,
which is used in the Standard Model. Whereas in quantum
field theories elementary particles are 
idealised point particles, the augmentation of the Standard Model
with gravity necessitates a departure from the view of elementary
particles as idealised points. We should not be surprised. There is 
nothing sacred about elementary particles as idealised points. 
The minimal hypothesis is to assume that elementary particles are 
not zero dimensional, but rather have one internal dimension, {\it i.e.}
strings. 
String theory provides a perturbatively self consistent framework for
quantum gravity. Furthermore, the consistency requirements of string
theory necessitate the appearance of the gauge and matter sectors 
that are the bedrock of the Standard Model. String theory therefore
provides a framework for the development of a phenomenological 
approach to quantum gravity. Nonperturbative extensions of string theory
reveal that in that context higher dimensional objects play a role as well.
However, to confront string theory with observational data, we may use any of its 
perturbative limits. 

Phenomenological string models have been constructed since the
mid--eighties. However, the majority of these constructions merely contain 
some of the ingredients of the Standard Model, such as possessing 
chiral families that are charged under some GUT gauge group, 
but do not offer room for more in depth analysis. 
A class of string models that exhibit realistic phenomenological properties,
and provide room for more in depth analysis, are the quasi--realistic models 
constructed in tne free fermionic formulation. 
These models correspond to $Z_2\times Z_2$ orbifolds of six dimensional
toroidal compactifications, and can be studied in any of the pertubative 
string limits, as phenomenological models as well as in cosmological 
scenarios. 
It should be understood, however, that our current understanding of string theory
is rudimentary. In particular, we only truly understand string
theory in its static limits and understanding of its time--dependent 
dynamics is still very much lacking. In order to explore time dependent
string dynamics is facilitated by going to the effective field theory
limit of the string constructions. Typically this effective field theory
limit only involves the massless degrees of freedom of the string models. 

This track has led to the "so--called" "Swampland Program". 
The aim of the Swampland Program is to address the question: when 
does an effective field theory model of quantum gravity have an ultraviolet complete
embedding in string theory. This approach can be viewed as a bottom--up
approach to the phenomenological exploration of string quantum gravity. 
An alternative top--down approach was advocated in this paper. The top--down
approach aims to explore the imprint of the string theory dualities and symmetries 
in the effective field theory limit. The most celebrated example of this 
approach is mirror symmetry. Mirror symmetry was first observed in worldsheet
construction of string vacua. It was entirely unexpected from the effective 
field theory point of view, and its profound implications in this context
were astounding. Spinor--vector duality, as described here, is an extension 
of mirror symmetry. While mirror symmetry correspond to mappings of the 
internal moduli of the compactified space, spinor--vector duality arises from 
mappings of the Wilson line moduli, and it provides a probe to explore the 
moduli spaces of $(2,0)$ string compactifications.

It is important to note that a notable characteristic of the top--down approach
is that it has access to the massive string modes, which are not seen 
in the effective field limit. Many of the string dualities are generated 
by the exchange of massless and massive string states, and are therefore 
naturally gleaned in the top--down approach, but are completely obscured 
in the effective field theory limit. Moreover, mirror symmetry and the spinor--vector
duality are mere two examples. String theory possess the nonperturbative dualities 
that generated some interest in the 1990s, but more importantly there are many dualities
that are yet to be explored and understood. An example of such a map that was discussed
here is the \ds--map, which induces a map between supersymmetric and non--supersymmetric
vacua in four dimensions. Understanding such mappings in depth is of vital importance 
because it may have implications for the dynamics of string theory, {\it i.e.}
it is a mapping between stable and unstable configurations. Even so, the space of
string symmetries to be explored is vast, and while the phenomenological models that can 
be constructed in string theory do suggest that it is relevant to the observed experimental data, 
our understanding of it is rudimentary. The field is still in its infancy. It is important, however, 
to discern what are the important questions to ask. In this respect 
the important question is not whether this or that string vacuum is the 
correct model of the world, but rather whether any of the properties of
the string vacua is relevant to the observable data. We know for certain 
that some of these properties are indeed relevant, {\it e.g.} the replication
of the fermion families. The aim is to go deeper in associating the properties 
of the string vacua with the experimentally observed data, {\it e.g.} in the 
flavour data. 

The next step in trying to formulate the theory of quantum gravity is to 
hypothesise some fundamental principles that may underlie quantum gravity 
and use them as a starting point to formulate the mathematical approach.  
I proposed here that $T$--duality may be interpreted as phase--space
duality in compact space. Manifest phase--space duality is the starting
point of the derivation of quantum gravity from phase--space duality
and the equivalence postulate of quantum mechanics. It should be opined that
the current string based approaches to the fundamental formulation of quantum
gravity are not satisfactory because they are background dependent. 
Phase--space duality and the equivalence postulate of quantum mechanics
are background independent principles. I proposed here that they provide the
overarching principles that underlie quantum gravity. The next steps in this saga 
are yet to be written and will occupy us in the millennia to come.

%%%%%%%%%%%%%%%%%%%%%%%%%%%%%%%%%%%%%%%%%%
%%%%%%%%%%%%%%%%%%%%%%%%%%%%%%%%%%%%%%%%%%

\bigskip

\section*{Acknowledgments}

\noindent
This work is supported in part by a Weston visiting professorship at the
Weizmann Institute of Science. 
I would like to thank Doron Gepner for discussions
and the Department of Particle Physics and Astrophysics for hospitality.

\bigskip
%\newpage

\bibliographystyle{unsrt}

\begin{thebibliography}{50}

% Reference 1
\bibitem{stringtheory} {\it For recent reviews, see e.g.:}\\
Mohaupt, T. \textit{A Short Introduction to String Theory}, 1st ed.; Cambridge University Press, 2022;\\
Kiritsis, E. \textit{String Theory in a Nutshell}, 2nd ed.; Princeton University Press, 2019.

\bibitem{loopy} {\it For review and references, see e.g.:}\\
Rovelli, C.; Vidotto, F. \textit{Covariant Loop Quantum Gravity: An Elementary Introduction to Quantum Gravity and Spinfoam Theory}, 1st ed.; Cambridge University Press, 2015.

\bibitem{mirrorsymmetry}  Greene, B.; R. Plesser, \textit{Duality in Calabi-Yau moduli space}, \NPB{338}{1990}{15};\\
  Candelas, P.; Lynker, M.; Schimmrigk, R. \textit{Calabi-Yau Manifolds in Weighted P(4)}, 
    \NPB{359}{1991}{21}.

\bibitem{mirrorreview} {\it For review and references, see e.g.:}\\
Hori, K.; \etal, 
\textit{Mirror Symmetry}, Clay mathematics monographs, American Mathematical Society, 2003.
    
\bibitem{dhvw} Dixon, L.J.; Harvey, J.A.; Vafa, C.; and Witten, E. 
\textit{Strings on Orbidolds},
  \NPB{261}{1985}{678}; \NPB{274}{1986}{285}.

\bibitem{fff}
Antoniadis, I.; Bachas, C.; Kounnas, C. 
\textit{Four-Dimensional Superstrings},
\NPB{289}{1987}{87};\\
Kawai, H.; Lewellen, D.C.; Tye, S.H.-H. 
\textit{Construction of Fermionic String Models in Four-Dimensions},
\NPB{288}{1987}{1};\\
Antoniadis, I.; Bachas, C. 
\textit{4-D Fermionic Superstrings with Arbitrary Twists},
\NPB{298}{1988}{586}.

\bibitem{gepner} Gepner, D. 
\textit{Exactly Solvable String Compactifications on Manifolds of SU(N) Holonomy},
  \PLB{199}{1987}{380}.

\bibitem{candelas} Candelas, P.; Horowitz, G.T. ; Strominger, A.; Witten, E.
\textit{Vacuum Configurations for Superstrings},
  \NPB{258}{1985}{46}.

\bibitem{IbUranga} For further review and references see {\it e.g.}:\\
  Ibanez, L.E.; Uranga, A.M.  \textit{An introduction to string phenomenology},
  Cambridge University Press, 2012.  

\bibitem{swamppro}
C.~Vafa,
\textit{The String landscape and the swampland},
[arXiv:hep-th/0509212 [hep-th]].

\bibitem{swamprev} {\it For recent review, see e.g.:}
Palti, E. \textit{The Swampland: Introduction and Review},
Fortsch. Phys. \textbf{67} (2019) 1900037
doi:10.1002/prop.201900037
[arXiv:1903.06239 [hep-th]].

\bibitem{DKL} Dixon, L.J.; Kaplunovsky, V.; Louis, J. \textit{On Effective Field Theories Describing (2,2) Vacua of the Heterotic String}, \NPB{329}{1990}{27}.

\bibitem{canxen}  Candelas, P.; de la Ossa, X.; Green P.; Parkes, L. \textit{A Pair of Calabi-Yau manifolds as an exactly soluble},                   \NPB{359}{1991}{21}.

\bibitem{syz} Strominger, A.; Yau, S.T.; Zaslow, E. \textit{Mirror symmetry is T duality}, \NPB{479}{1996}{243}.

\bibitem{gprreview}  Giveon, A.; M. Porrati, M.; Rabinovici, E. \textit{Target space duality in string theory}, \PRT{244}{1994}{77}.

\bibitem{svd1} \AEF; Kounnas, C.; Rizos, J. \textit{Chiral family classification of fermionic $Z_2\times Z_2$ heterotic orbifold models}, \PLB{648}{2007}{84}. 

\bibitem{svd2} \AEF; Kounnas, C.; Rizos, J. \textit{Spinor-Vector Duality in fermionic $Z_2\times Z_2$ heterotic orbifold models},
\NPB{774}{2007}{208};

\bibitem{ffmt} \AEF; Florakis, I.; Mohaupt, T.; Tsulaia, T. \textit{Conformal Aspects of Spinor-Vector Duality}, \NPB{848}{2011}{332}.

\bibitem{afg} Athanasoupoulos, P.; \AEF; Gepner, D. \textit{Spectral flow as a map between $N = (2,0)$-models}, \PLB{735}{2014}{357}.

\bibitem{fkrneq2} \AEF; Kounnas, C.; Rizos, J. \textit{Spinor-vector duality in $N=2$ heterotic string vacua}, \NPB{799}{2008}{19}.

\bibitem{5dsvd} \AEF; Groot--Nibbelink, S.; Hurtado--Heredia, M.; \textit{Uncovering a spinor–vector duality on a resolved orbifold}, \NPB{969}{2021}{115473}. 

\bibitem{6dsvd} \AEF; Groot--Nibbelink, S.; Hurtado--Heredia, M.; \textit{Constraint on spinor-vector dualities in six dimensions}, \PRD{103}{2021}{126016}.

\bibitem{taming} \AEF; Groot--Nibbelink, S.; Hurtado--Heredia, M.; 
\textit{Taming triangulation dependence of $T^{6}/Z_{2}\times Z_{2}$ resolutions}, 
\JHEP{01}{2022}{169}.

\bibitem{gkr} Gregori, A.; Kounnas, C.; Rizos, J. \textit{Classification of the $N=2$, $Z_2\times Z_2$ symmetric type II orbifolds and their type II asymmetric duals}, \NPB{549}{1999}{16}.

\bibitem{fknr} \AEF; Kounnas, C.; S.E.M Nooij; S.E.M.; Rizos, J. \textit{Classification of the chiral 
$Z_2\times Z_2$ fermionic models in the heterotic superstring}, \NPB{695}{2004}{41}.

\bibitem{cfkr}
  Catelin-Julian, T.; \AEF; Kounnas, C.; Rizos, J.; 
  \textit{Spinor--Vector Duality in Heterotic SUSY Vacua}
                                               \NPB{812}{2009}{103}.
\bibitem{aft} Angelantonj, C.; \AEF; Tsulaia, M.; 
\textit{Spinor-Vector Duality in Heterotic String Orbifolds}, 
\JHEP{07}{2010}{004}.

\bibitem{angelsign}  Angelantonj, C.; Sagnotti, A.; 
\textit{Open Strings}, 
\PRT{371}{2002}{1}.
                           
\bibitem{msds} Florakis, I.; Kounnas, C. 
\textit{Orbifold Symmetry Reductions of Massive Boson-Fermion Degeneracy}, \NPB{820}{2009}{237}; 
\textit{Marginal Deformations of Vacua with Massive boson-fermion Degeneracy Symmetry}, \NPB{834}{2010}{273}.

\bibitem{panos} P. Athanasopoulos and \AEF, 
\textit{Niemeier Lattices in the Free Fermionic Heterotic–String Formulation},
                {\it Adv.\ Math.\ Phys.}\/ {\bf 2017} (2017) 3572469. 
                
\bibitem{vafawitten} Vafa, C.; Witten E.;
\textit{On orbifolds with discrete torsion}, 
    {\it J.Geom.Phys.} {\bf 15} (1995) 189. 
                
\bibitem{nonsusy} \AEF; Matyas, V.G.; Percival, B.; 
\textit{Towards the Classification of Tachyon-Free Models From Tachyonic Ten-Dimensional Heterotic String Vacua}, 
\NPB{961}{2020}{115231}; 
\textit{Classification of nonsupersymmetric Pati-Salam heterotic string models}, 
\PRD{104}{2021}{046002};
\textit{Towards Classification of $\mathcal{N}=1$ and $\mathcal{N}=0$ Flipped $SU(5)$ Asymmetric $\mathbb{Z}_2 \times \mathbb{Z}_2$ Heterotic String Orbifolds},
\PRD{106}{2022}{026011}, arXiv:2202.04507.


\bibitem{nahe} \AEF; Nanopoulos, D.V.;
\textit{Naturalness of three generations in free fermionic $Z_2^n \times Z_4$ 
string models}, 
\PRD{48}{1993}{3288}.

\bibitem{spwsp} \AEF; 
\textit{String Phenomenology from a worldsheet perspective}, 
\EJP{79}{2019}{703}.

\bibitem{stable} \AEF; Matyas, V.B.; Percival, P.; 
\textit{Stable Three Generation Standard--like Model From a Tachyonic Ten Dimensional Heterotic--String Vacuum}, 
\EJP{80}{2020}{337}.

\bibitem{ginspargvafa} Ginsparg, P.H.; Vafa, C.;
\textit{Toroidal Compactification of Nonsupersymmetric Heterotic Strings}, 
\NPB{289}{1987}{414}.

\bibitem{itoyamataylor} Itoyama, H.; Taylor, T.R.;
\textit{Supersymmetry Restoration in the Compactified 
$O(16)\otimes O(16)^\prime$ Heterotic String Theory}, 
\PLB{186}{1987}{129}.

\bibitem{ginsparg} Ginnsparg, P.; 
\textit{Comment on Toroidal Compactification of Heterotic Superstrings}, 
\PRD{35}{1987}{648}.

\bibitem{hull} Hull, C.M.; 
\textit{A Geometry for non-geometric string backgrounds}, 
\JHEP{10}{2005}{065}.

\bibitem{fm2} Faraggi, A.E.; Matone, M.; 
\textit{Quantum mechanics from an equivalence principle}, 
\PLB{450}{1999}{34}. 

\bibitem{epoqm}  Faraggi, A.E.; Matone, M.; 
\textit{The Equivalence postulate of quantum mechanics}, 
\IJMP{15}{2000}{1869}. 

\bibitem{bfm}  Bertoldi, G.; Faraggi, A.E.; Matone, M.; 
\textit{Equivalence principle, higher dimensional Mobius group and the hidden antisymmetric tensor of quantum mechanics}, 
 {\it Class.\ Quant.\ Grav.}\/ {\bf 17} (2000) 3965.

\bibitem{floyd} Floyd, E.R.;
\textit{Bohr-Sommerfeld quantization with the effective action variable},
\PRD{25}{1982}{1547}; 
\textit{Modified potential and Bohm's quantum mechanical potential},
{\bf D26} (1982) 1339; 
\textit{Arbitrary initial conditions of nonlocal hidden variables},
{\bf D29} (1984) 1842;
\textit{Closed-form solutions for the modified potential},
{\bf D34} (1986) 3246;
\textit{Reflection time and the Goos-H\"anchen effect for reflection by a semi-infinite rectangular barrier},
   {\it Found.\ Phys.\ Lett.}\/ {\bf 13} (2000) 235;
\textit{Classical limit of the trajectory representation of quantum mechanics and residual indeterminacy}, 
 \IJMP{15}{2000}{1363}.

\bibitem{holland} Holland, P.R.;
\textit{The de Broglie-Bohm theory of motion and quantum field theory},
\PRT{224}{1993}{95}. 

\bibitem{gravcon} \AEF~and M. Matone, 
\textit{The Geometrical Origin of Dark Energy}, 
\EJP{80}{2020}{1094}. 





\end{thebibliography}

\end{document}